\documentclass[12pt]{article}
\usepackage{latexsym}
\usepackage{amssymb}
\usepackage{graphicx,epsfig}
\usepackage{rotating}
\usepackage{relsize}
\usepackage{graphicx}
\usepackage{amsfonts}
\usepackage{amsmath}
\usepackage{booktabs}
\RequirePackage[round,sort&compress]{natbib}
\RequirePackage[colorlinks,citecolor=blue,urlcolor=blue,linkcolor=blue]{hyperref}
\usepackage{authblk}
\makeatletter
\newcommand{\singlespacing}{\let\CS=\@currsize\renewcommand{\baselinestretch}{1}\tiny\CS}
\newcommand{\oneandahalfspacing}{\let\CS=\@currsize\renewcommand{\baselinestretch}{1.25}\tiny\CS}
\newcommand{\doublespacing}{\let\CS=\@currsize\renewcommand{\baselinestretch}{1.35}\tiny\CS}

\def\@citex[#1]#2{\if@filesw\immediate\write\@auxout{\string\citation{#2}}\fi
  \def\@citea{}\@cite{\@for\@citeb:=#2\do
    {\@citea\def\@citea{,\linebreak[0]\hskip0pt plus .2em}%
      \@ifundefined{b@\@citeb}%
    {{\bf ?}\@warning{Citation `\@citeb' on page \thepage\space undefined}}%
      \hbox{\csname b@\@citeb\endcsname}}}{#1}}

\date{}
\RequirePackage{graphicx}
\usepackage{float}
\RequirePackage[round,sort&compress]{natbib}
\RequirePackage[colorlinks,citecolor=blue,urlcolor=blue,linkcolor=blue]{hyperref}
\begin{document}
\title{\bf The classification of interior solutions of anisotropic fluid configurations}  
\author{\bf Jitendra Kumar\thanks{\color{blue}jitendark@gmail.com} }
\author{\bf Puja Bharti \thanks{\color{blue}pujabharti06@gmail.com} }
\affil{\small Department of Mathematics, Central University of Jharkhand, Cheri-Manatu, Ranchi-835222 India.}

\date{}
\maketitle \setlength{\parskip}{.11in}
\setlength{\baselineskip}{15pt}

\begin{abstract}
The Einstein-Maxwell (or Einstein) system of field equations plays a substantial role in the modeling of compact stars. Although due to its non-linearity getting an exact solution for the system of field equations is a difficult task, the solutions of field equations have a long and rich history. It took a year for Karl Schwarzschild to obtain the first exact solution of Einstein’s field equations since general theory of relativity was published. The number of viable solutions has been growing since then. Many authors have adopted several methods to obtain the solution. Different models have been constructed for a variety of applications. 
To produce feasible models of compact stars, a considerable amount of effort has been applied in gaining an understanding of the properties of anisotropic matter. Theoretical study indicates that pressure within compact stars with extreme internal density and strong gravity is mostly anisotropic. Anisotropy was found sufficient for the study of compact stars with the dense nuclear matter. It is claimed that it is important to consider the pressure experienced to be anisotropic whenever relativistic fluids are involved. In this review article, we have discussed different ways of generating a static spherically symmetric anisotropic fluid model. The purpose of the article is to present a simple classification scheme for static and spherically symmetric anisotropic fluid solutions. The known solutions are reviewed and compartmentalized as per the proposed scheme so that we can illustrate general ideas about these solutions without being exhaustive.
\end{abstract}

\maketitle

\section{Introduction}\label{sec1}
\subsection{Compact Stars}	
Stars are formed in gas and dust clouds with a non-uniform matter distribution and scattered throughout most galaxies. In the process of their evolution, all active stars eventually come to a point, when the outward radiation pressure from the nuclear fusions in its interior no longer resists the ever-present gravitational forces. This is when the star collapses due to its own weight and it undergoes the process of stellar death \citep{Sagert}. For most stars, this is the process that is responsible for the formation of a very dense and compact stellar remnant, known as a compact star. In other words, Compact stars, which includes white dwarfs, neutron stars, black holes, and quark star, are the final stages in the evolution of ordinary stars. The phase separations of the early universe following the big bang can also be responsible for the formation of compact stars. Apart from having a very high density, compact stars are characterized by the fact that nuclear reactions completely cease to exist in their interiors. This is the reason why they cannot support themselves against gravity. In the white dwarfs and neutron stars, gravity faces resistance by the pressure of degenerated gas. In black holes, the force of gravity is completely dominant over other forces and compresses the stellar material resulting in infinite density \citep{Karttunen}.
When a star like the Sun completely exhausts its nuclear fuel, its core collapses to form a dense white dwarf. But, when inert iron cores of stars having mass equal to or greater than ten times that of the Sun collapse then it forms an extremely dense neutron star or black hole. Although the universe is not that old for any of the smallest red dwarfs to have reached the end of their existence, stellar models suggest that they will slowly become brighter and hotter before running out of hydrogen fuel and eventually it will become low-mass white dwarfs.

Depending on the mass during the lifetime of a star, when a star ceases its nuclear fuel supply, its remnants can take one of these forms :
\begin{enumerate}
	\item White and black dwarfs: ``White or degenerate dwarfs" mainly consists of degenerated matter; typically carbon and oxygen nuclei in a sea of degenerated electrons. White dwarfs are formed from the core of main-sequence stars and therefore they are very hot initially when they are formed. As they cool they will get dimmer and redden until they eventually become dark and transform into ``black dwarfs". However, our universe has not aged enough for any black dwarfs to exist yet. White dwarfs are stable because there is a balance between the inward pull of gravity and the degeneracy pressure of the star's electrons.
	After the degenerate star's mass grows sufficiently, i.e., when its radius shrinks to only a few thousand kilometers, the mass will approach the Chandrasekhar limit (about 1.4$M_\odot$), which is the upper limit (theoretical) of the mass of a stable white dwarf.
	\item Neutron stars: If the white dwarf's mass increases above the Chandrasekhar limit, then electron degeneracy pressure fails due to electron capture and the star collapses into a neutron star. In case of certain binary stars such as white dwarf star, mass is transferred from the companion star onto the white dwarf star, eventually pushing it over the Chandrasekhar limit. Electrons react with protons to create neutrons and thus the required pressure is not longer supplied to resist gravity, causing the star to collapse. With a further increase in density, the remaining electrons react with the protons to form more neutrons. It continues to collapse until the neutrons become degenerate. A new equilibrium is possible after the star shrinks by three orders of magnitude, to reach a radius between 10 and 20 km. This is a ``neutron star". These stars are extremely small and due to the conservation of angular momentum, their period of rotation shortens as the stars shrink and it ranges from about 1.5 milliseconds to several seconds. When magnetic poles of these rapidly rotating stars are aligned with the Earth, a pulse of radiation can be detected in each revolution. Such neutron stars are called pulsars.
	\item Black holes: If the mass of the stellar remnant is high enough, the neutron degeneracy pressure will be insufficient to prevent collapse below the Schwarzschild radius. As a consequence, the stellar remnant becomes a ``black hole". With the accumulation of more mass, equilibrium against  gravitational collapse  exceeds its breaking point. Once the star's pressure is not sufficient enough to counterbalance gravity, a catastrophic implosion occurs within a fraction of milliseconds. The escape velocity at the surface quickly reaches the velocity of light. At that point a black hole gets formed as no energy or matter can escape. The mass at which this occurs is not known with certainty but is currently estimated at between 2 and 3 $M_\odot$.
	\item Quark stars or strange stars: If neutrons are squeezed sufficiently at a very high temperature, they will decompose into their component quarks, forming a quark matter. This will shrink the star further and make it denser, but instead of a total collapse into a black hole, the star may stabilize itself and survive in this state indefinitely, as long as no more mass is added further. It will, to an extent, become a very large nucleon. A star in this hypothetical state is known as a ``quark star" or more specifically a ``strange star".
\end{enumerate}	
\subsection{History of Compact Stars}			 
Since around the mid-1920s, physicians started using stars as “physics laboratories” for investigating how matter changes its nature under extreme densities and pressures, which was impossible to realize on Earth. The details on the process of evolution of the concept of compact stars can be found in \citep{Luisa}. Robert Oppenheimer was the first person who started the discussion on the possibility of gravitational collapse. His contribution can also be found in the first systematic application of Einstein’s general theory of relativity to a compact star. As white dwarfs had provided in 1915 a new test of Einstein’s theory of general relativity \citep{Connor} outside the solar system, the neutron star hypothesis, in conjunction with observations on supernovae lead to a further and far-reaching test of the general theory of relativity.

In 1924 Arthur S. Eddington discussed white dwarfs in detail. Also, Frenkel classified the super-dense stars into two categories, one consisting of non-relativistic and the other consisting of the ultra-relativistic electron gas. He correctly estimated the mass of a stable star, which is in a relativistic degenerate state and cannot exceed a definite maximum $ M (\ge M_{\odot} )$. To govern the structure of a star in radiative equilibrium, some fundamental equations have been established. In 1926, a British physicist Ralph H. Fowler examined the problem of degenerate dense matter in white dwarf stars \citep{ Fowler}. At the end of the 1930s, Oppenheimer contributed to the relativistic gravitational collapse of a neutron star by referring to different theoretical frameworks. These works marked the beginning of further investigations on the connection between compact objects and general relativity.

In 1926, Ralph H. Fowler also found that the relationship between the density, energy, and temperature of white dwarfs could be determined by considering them as a gas of non-relativistic, non-interacting nuclei that obey Fermi–Dirac statistics \citep{Fowler,Fowler2}. This Fermi gas model was brought into practice to calculate the relationship among the mass, radius, and density of the white dwarf stars, assuming them to be essentially homogeneous spheres of electron gas \citep{Stoner}. Wilhelm Anderson applied a relativistic correction to this model and in December 1928 put forward a theory according to which “the mass of a star must have a maximum value” \citep{Anderson}. At the end of 1929, Milne presented his investigations on the relation between the masses, luminosities, and effective temperatures of the stars \citep{Milne}.
In 1930, Stoner derived the internal energy–density equation of state for Fermi gases. With the help of this, he was finally able to treat the mass-radius relationship in a fully relativistic manner and concluded that it has a limiting mass of approximately $2.19\times 10^{30}$ kg \citep {StonerC}.
A series of papers \citep{Chandra1, Chandra2, Chandra3, Chandra4, Chandra5, Chandra6, Chandra7, Chandra8, Chandra9} were published between 1931 and 1935 by Chandrasekhar. In these papers, he solved the hydrostatic equation together with the nonrelativistic Fermi gas equation of state. He also contributed towards the case of a relativistic Fermi gas, giving rise to the value of the initial pre-defined limit. This new limit is known as the Chandrasekhar limit. 
The currently accepted value of the Chandrasekhar limit is about 1.4 solar mass \citep{Mazzali}.

During summer 1938, Zwicky came into the picture with a follow-up of his proposal about the existence of neutron stars \citep{Baade1, Baade2} and made a new attempt along this path employing general relativity. He made a stronger connection to general relativity, according to which “the mass of a star of given density cannot surpass a certain critical value (Schwarzschild limit)” \citep{Zwicky1}. He discussed the Schwarzschild solution\citep{Schwarzschild} and suggested the use of Schwarzschild’s interior solution for a sphere of fluid of constant density as providing a model for a “collapsed neutron star”. As per his conclusion, any star, when it reaches the Schwarzschild limiting configuration then it must be regarded as an object between which and the rest of the universe practically no physical communication is possible, therefore, it is not at all possible to observe any physical conditions in stellar bodies which have reached the Schwarzschild limit"\citep{Zwicky2}.

The redshift effect, which had been mentioned by Ernest J. Opik in his theory of giant stars as a phenomenon, may asymptotically tend to reduce the luminosity of a super-dense contracting star to zero \citep{Opic}.
In 1938 Oppenheimer and Serber, were able to do a reasonable estimation of the minimum mass for which such a core would be stable (approximately $0.1  M_\odot$) \citep{Serber} by taking into account some effects of nuclear forces. To differentiate between the rest mass and the gravitational mass, Zwicky had estimated the binding energy of a neutron star of mass M, and thus he was able to evaluate the quantity of energy that could be released during the core collapse of massive stars. Zwicky, based his theoretical investigations on spectral studies of two bright supernovae (IC 4182 and NGC 1003) performed by Rudolf Minkowski \citep{Minkowski}, which fully justified a more detailed examination of the neutron star as per him. But it was again Richard C. Tolman who inspired investigations towards the application of the general theory of gravitation. J. Robert Oppenheimer and George Volkoff applied general relativity to tackle the problem of a compact astrophysical object for the first time. In 1939, they were able to demonstrate that the general relativistic field equations do not guarantee any static solution for a spherical distribution of cold neutrons, if the total mass of the neutrons is greater than approximately $0.7  M_\odot$, and had established that a star under these circumstances would collapse under the influence of its gravitational field \citep{Oppenheimer}. This limiting value for cold and non-rotating neutron star mass is known as the Tolman–Oppenheimer–Volkoff limit. This limit was analogous to the Chandrasekhar limit for the mass of white dwarf stars. Taking account of the strong nuclear repulsion forces between neutrons, modern work leads to estimates of this limit in the range from approximately $1.5-3 M_\odot$ \citep{Bombaci}. However, it was noticed that the maximum supportable mass can be increased by rotation \citep{Friedman, Cook, Eriguchi, Paschalidis}.

Snyder and Oppenheimer had explored the process of gravitational collapse. In their work, the full consequences of Einstein’s theory of gravitation are there \citep{Snyder}. By 1939, a handful of researchers had already worked on the problem of what happens to a compact star core made entirely of degenerate fermions (electrons and neutrons). Because of these contributions, on one side nuclear matter and particle physics were becoming essential for the description of matter at such extreme densities and on the other hand, it had become clear that such super-dense objects could be described only within Einstein’s theory of gravity. The door opened for the world of relativistic astrophysics.

\subsection{Anisotropy and its effect}
The interior Schwarzschild solution, from which all problems involving spherical symmetry can be modeled, represents the starting point in the study of fluid spheres.
Theoretical relativistic stellar models are wont to predict many crucial properties of compact objects. To get realistic models of compact objects, a substantial amount of effort has been applied in gaining an understanding of the properties of anisotropic matter.
Generally, compact stars are considered spherically symmetric and isotropic. However, isotropy is not a general characteristic of stellar objects. Extreme internal density and strong gravity of compact objects hint that pressure within these objects might not be in the form of an ideal fluid. Theoretical study indicates that pressure within such stars is usually anisotropic. This suggests that the pressure inside the fluid sphere can specifically be decomposed into two non-identical parts: the radial pressure, $p_r$ and the transverse pressure, $p_t$, which acts in a mutually perpendicular direction. Anisotropic factor $(\Delta = p_t-p_r)$, which deals with the inhomogeneity in these two pressures, is a function to examine the interior situation of the star compared to the idealized isotropic case. A scalar field with a non-zero spatial gradient is a physical system with anisotropic pressure. The so-called ‘boson stars’ are naturally anisotropic \citep{Schunck2003}. Similarly, the energy-momentum tensor of both electromagnetic and fermionic fields are naturally anisotropic \citep{Dev}. Wormholes and gravastars are also considered anisotropic in nature.

The idea that the tangential pressure may be different from the radial pressure was suggested first by Lemaitre in 1933 \citep{Lemaitre}. This model was entirely dependent on tangential pressure and has constant density throughout. The model was generalized further for variable density by Florides \citep{Florides}. But, the concept of anisotropy was originally proposed by Ruderman \citep{Ruderman} and later on several other scientists put their effort into it. 
As the reasons behind the existence of anisotropy in the star model, different factors are found to be responsible. It is claimed that anisotropy may develop inside the stellar objects in the high-density regime ($> 10^{15} g/cc$) of compact stars, where the nuclear interactions must be treated relativistically \citep{Ruderman,Canuto}. 
It is also argued that anisotropy in pressure may arise due to the formation of super-fluid neutrons inside stars \citep{Kippenhahn}. There is several research work present in literature to validate the fact that pressure anisotropy can be triggered by the presence of solid core, various types of phase transition \citep{Sokolov}, pion-condensation \citep{Sawyer}, slow rotation \citep{Herrera95}, strong magnetic field \citep{Weber}, a mixture of two fluids \citep{Letelier, Bayin} etc. 
Viscosity may be an additional source of local anisotropy. Some numerical calculations on the effects of viscosity-induced anisotropy are presented by Barreto et al. \citep{Barreto92,Barreto93}. A comprehensive list of physical phenomena producing pressure anisotropy and a review on effects of local anisotropy in astrophysical objects can be found in \citep{Herrera,Chan1997}. The factors contributing to the pressure anisotropy have also been discussed in \citep{Dev2002,Dev2003}. However, in the later stage, to understand its nature and consequences, diversified investigations have been performed on ultra-dense spherically symmetric fluid spheres having pressure anisotropy. Ivanov \citep{Ivanov2010} mentioned that if we consider the system to be anisotropic then influences of shear, electromagnetic field, etc. on self-bound systems can be absorbed. Recently, it is concluded by Herrera \citep{Herrera20} that even if the system is initially assumed to be isotropic, physical processes are expected in stellar evolution like dissipative fluxes, energy density inhomogeneities, or the appearance of shear in the fluid flow, will always tend to produce pressure anisotropy. As an equilibrium configuration is the final stage of a dynamic regime, the acquired anisotropy during this dynamic process won't disappear in the final equilibrium state, and thus the resulting configuration, even though initially had isotropic pressure, should in principle exhibit pressure anisotropy. This is why it is important to consider pressure anisotropy whenever relativistic fluids are involved.

Bowers and Liang \citep{Bowers} was first to apply the anisotropic model on the equilibrium configuration of relativistic compact star-like neutron stars. They also claimed that anisotropy might have non-negligible effects on surface redshift and equilibrium mass. Several anisotropic models have been developed by incorporating anisotropic pressure in the stress-energy tensor of the material composition. The application of anisotropic fluid models to neutron stars was done both analytically and numerically \citep{Heintz,Steinmetz,Cosenza, Stewart, Bayin, Krori, Maharaj, Bondi, Gokhroo, Patel}. In these works different exact solutions of the Einstein field equations for anisotropic spheres had been obtained which can be used as models of massive compact objects. Heinzmann and Hillebrandt \citep{Heintz} have considered fully relativistic anisotropic superdense neutron star models and have established that there is no limiting mass of a neutron star for arbitrarily large anisotropy. However, the maximum mass of the neutron star was still lying beyond $3-4M_{\odot}$. Hillebrandt and Steinmetz\citep{Steinmetz} have discussed the stability of fully relativistic anisotropic neutron stars and found that there exists a static stability criterion similar to the one obtained for isotropic models.  
Maharaj and Maartens \citep{Martens} discussed a solution for an anisotropic fluid sphere with uniform density, whereas Patel and Mehta \citep{Patel} did the same for the case of variable density distribution. For static spheres in which the tangential pressure differs from the radial one, Bondi \citep{Bondi} has studied the link between the potential at the surface and the highest occurring ratio of the pressure tensor to the local density. Utilizing the Maharaj and Maartens \citep{Martens} algorithm, Gokhroo and Mehra \citep{Gokhroo} and Chaisi and Maharaj \citep{Chaisi05,Chaisi06} have developed and studied new anisotropic fluid models. Herrera and Santos \citep{Herrera95} have extended the Jeans instability criterion in Newtonian gravity to systems with anisotropic pressures. Dev and Gleiser \citep{Dev2002} obtained several new exact solutions for stars of constant density and to each of the solutions they had demonstrated that anisotropy may indeed change the general properties of the stars such as critical mass and surface redshift of the equilibrium configurations.  Ivanov \citep{Ivanov} has investigated the surface redshift of anisotropic realistic stars and shown that for realistic anisotropic star models the surface redshift can not exceed the value 3.842 when the tangential pressure satisfies the strong energy condition and the value 5.211 when the tangential pressure satisfies the dominant energy condition. Chan et al. \citep{Chan1997} studied in detail the role played by the local pressure anisotropy in the onset of instabilities and they showed that small anisotropies might in principle drastically change the stability of the system. General algorithms for generating static anisotropic solutions were also found by Lake \citep{Lake09}.

It was  Maurya  \citep{Mourya18}  who argued that the interaction among the particles is too relativistic and they become too random to maintain any uniform distribution throughout the region. Based on this argument he suggested that the chances of having anisotropy are much higher in compact stars. It is found that the relativistic nature of particles in compact stars can generate a significant anisotropy in the compact star. As a result, the anisotropic force inside the star is claimed to be responsible for making the compact object more compact in comparison to the isotropic condition, which eventually makes the possible transition of a neutron star to a strange star. Later, Jasim \citep{Jasim}  presented a model of an anisotropic fluid sphere under Einstein’s general theory of relativity by employing the Tolman-Kuchowicz metric \citep{Tolman39,Kuchowicz}. Thomas and coresearchers \citep{Thomas05,Thomas07,Tikekar05} have proposed models of gravitationally bound systems in equilibrium with an anisotropic fluid distribution. Some other anisotropic models corresponding to static spherically symmetric anisotropic matter distributions can be found in  \citep{Bondi2,Barreto92,Barreto93,Coley,Mart,Singh,Hern,Dev2002,Dev2003,Harkomak2000,Lake,Bohmer1,Bohmer2,Esculpi07,Khadekar,Karmakar07,Abreu,Ivanov2010,Herrera081,Mak2003,Sharma,Harkomak,Herrera08,Maurya19,amit19}. Very recently Kumar and Bharti \citep{Bharti} has obtained an  anisotropic fluid model by employing the Vaidya-Tikekar \citep{Vaidya} metric potenntial. 

Generating fluid model by employing equation of state has also become popular. Assuming a linear EoS, Maharaj and Chaisi \citep{Maha2006}, Sharma and Maharaj \citep{Sharma07}, Esculpi and Aloma \citep{Esculpi10} provided exact analytic solutions for the compact anisotropic matter distributions. Recently, Malaver \citep{Malaver18} have also analyzed an anisotropic fluid distribution to obtain a new class of exact solutions. Using the Finch and Skea \citep{Finch} ansatz for the metric potential, Sharma and Ratanpal \citep{shar13} have reported a static spherically symmetric compact anisotropic star model which admits a quadratic EoS. Pandya et al. \citep{Pandya}, Bhar et al. \citep{Bhar359} studied the static spherically symmetric relativistic anisotropic compact object considering the Tolman VII solution as one of the metric potentials. Mardan et al. \citep{Mardan} has used polytropic EoS to generate solutions. New anisotropic models of a strange star admitting the Chaplygin equation of state was proposed by Bhar \citep{bhar15}, Kumar et al. \citep{amit}, etc.

Fuzfa et al. \citep{Fuzfa} did an analysis based on the Weyl tensor of the Lemaitre-Schwarzschild problem of finding the equilibrium conditions of an anisotropically sustained spherical body under its relativistic gravitational field. A general method for obtaining static and anisotropic spherically symmetric solutions satisfying a nonlocal equation of state from known density profiles, by assuming the condition of a vanishing Weyl tensor is presented by Hernandez and Nunez \citep{Hernandez}. This condition can be integrated for the spherically symmetric case. Later, the resulting expression is used by  Herrera et al. \citep{Prisco} to find a new model to see the effect of local anisotropy on the Tolman-Whittaker mass \citep{Tolman,Whittaker} distribution within the sphere, and on the velocity profile of different pieces of matter.  Some classes of exact solutions of Einstein field equations describing spherically and static anisotropic stellar type configurations were presented in \citep{Harkomak, Mak}. Mak et al. \citep{Mak} has presented the general solution of the gravitational field equations for an anisotropic static matter distribution assuming mass function in the functional form and a specific mathematical representation of the anisotropy parameter. In their model $\Delta$ has the same qualitative properties as the anisotropy parameter of boson stars. 

A large number of works on conformal motion have been published by several authors.  Rahaman et al. \citep{Rahman10} has studied a class of solutions for anisotropic stars admitting conformal motion.  Relativistic stars admitting conformal motion have been analyzed in \citep{Rahman102}.  A charged fluid sphere with a linear equation of state admitting conformal motion has been studied in \citep{Esculpi10}. Ray et al. \citep{Ray07} has given an electromagnetic mass model admitting a conformal Killing vector.  Mak and Harko \citep{Mak} have described a charged strange quark star model by considering a one-parameter group of conformal motion.  A charged gravastar admitting conformal motion has been studied by Usmani et al. \citep{Usmani}. Contrary to this work, Bhar \citep{bhar14} have studied a higher-dimensional charged gravastar admitting conformal motion. Rahaman et al. \citep{Rahman15} have also described conformal motion in higher-dimensional spacetime.  Bhar \citep{Bhar15} has obtained a new class of solutions of anisotropic stars under the framework of general relativity inspired by noncommutative geometry in four and higher-dimensional spacetime. Recent works on conformal motion can be found in \citep{Rahman 17, Matondo, Maurya2019}.

The supposition of local anisotropy of pressure has been found very useful in the study of relativistic compact objects. It is crystal clear from the expression of the anisotropy factor, that it could be both positive or negative. Both the cases have different significance. When the tangential pressure dominates the radial pressure then the positive anisotropy factor generates an outwardly acting force and the stellar system experiences a repulsive force, that counteracts the gravitational gradient. On the other hand, when the anisotropy factor is negative and hence the radial pressure dominates the tangential pressure, it generates an inward force and the system experiences an attractive force. Moreover, a positive anisotropy factor enhances the stability and equilibrium of the system \citep{Mourya18,Morales}.

\subsection{Spherically symmetric space-time: Metric and Field equations}
For modeling some simplifying assumptions are taken:  Compact stars are assumed to be spherical and symmetrical, all physical quantities depend only on radial coordinate $r$,  rotations and outside gravitational influences are generally ignored. It is considered that the initial composition is uniform i.e., no initial dependence of composition on the radius,  gravity is Newtonian and stars change slowly with time.

The interior space–time in cannonical comoving coordinates is described by  the following spherically symmetric line element 
\begin{equation}
	ds^2=e^{\nu(r)}dt^2-e^{\lambda(r)}dr^2-r^2 d\Omega^2
	\label{metric}
\end{equation}
where,  we have used $d\Omega^2$ for $(d\theta^2+sin^2\theta d\phi^2)$. The metric potentials $\nu(r)$ and $\lambda(r)$ are arbitrary functions of radial coordinate $r$. 
To investigate the solutions of Einstein’s field equations for anisotropic compact stars, the energy-momentum tensor is considered in the following form:
\begin{equation}
	T_{j}^{i} = (p_r-p_t)\chi^i\chi_j+(\rho+p_t)v^iv_j-p_t\delta_j^i+E_{j}^{i}
	\label{tensor}
\end{equation}
with $v^iv_i=-\chi^i\chi_i=1$ and $u^i\chi_i=0$, the vector $v^i$ is the fluid four-velocity and $\chi^i$ is the unit spacelike vector in the radial direction which is orthogonal to $v^i$, and $E_{j}^{i}$ is the electromagnetic energy tensor.
For the spherically symmetric metric of Eq. (\ref{metric}) the Einstein-Maxwell field equation provides the following relationship:
\begin{eqnarray}
	\frac{\lambda'}{r}e^{-\lambda}+\frac{1-e^{-\lambda}}{r^2}=c^2 \kappa \rho + \frac{q^2}{r^4} \label{fe1}\\
	\frac{\nu'}{r}e^{-\lambda}-\frac{1-e^{-\lambda}}{r^2}=\kappa p_r- \frac{q^2}{r^4} \label{fe2}\\
	\Big(\frac{\nu''}{2}-\frac{\lambda' \nu'}{4}+\frac{\nu'^2}{4}+\frac{\nu'-\lambda'}{2r}\Big)e^{-\lambda}=\kappa p_t+ \frac{q^2}{r^4} \label{fe3}
\end{eqnarray}
where $\rho$ is the matter density, $p_r$ is the radial pressure and $p_t$ is the tangential one. Here prime denotes differentiation with respect to $r$ and $\kappa= \frac{8\pi G}{c^4}$, where, $G$ stands for gravitational constant and $c$ is the speed of light. 

Equations (\ref{fe2},\ref{fe3}) give the expression for anisotropic factor as
\begin{eqnarray}
	\kappa \Delta=\kappa (p_t-p_r)=	
	\Big(\frac{\nu''}{2}-\frac{\lambda' \nu'}{4}+\frac{\nu'^2}{4}-\frac{\nu'+\lambda'}{2r}\Big)e^{-\lambda}+\frac{1-e^{-\lambda}}{r^2}-\frac{2q^2}{r^4} 
	\label{fe4}
\end{eqnarray}
When pressure of matter inside the sphere will be isotropic then we take $P_r = p_t = p$, i.e., a zero $\Delta$. For the case when the electric field is absent, the charge function $q$, which measures the charge within radius $r$, will be $0$. When we charge a neutral solution, it decreases its anisotropic factor.

The metric potential functions $\nu(r)$ and $\lambda(r)$ uniquely determine the surface redshift and gravitational mass function respectively. The gravitational mass ($m$) in a sphere of radius $r$ is given by
\begin{eqnarray}
	\frac{2m}{r}=1-e^{-\lambda}+\frac{q^2}{r^2}
\end{eqnarray}
and the gravitational redshift ($z$) is related to metric potential $\nu$ by the following relationship
\begin{eqnarray}
	z(r)=e^{-\nu/2}-1
\end{eqnarray}

For $r=R$ as the outer boundary of the fluid sphere, $M=m(R)$, and $Q=q(R)$, the unique exterior spacetime of the compact star can be described by the following Reissner-Nordstr\"{o}m metric
\begin{equation}
	ds^2 =  \Big(1-\frac{2M}{r}+\frac{Q^2}{r^2}\Big)dt^2 - \Big(1- \frac{2M}{r}+\frac{Q^2}{r^2}\Big)^{-1}dr^2-r^2\big(d\theta^2 + sin^2\theta d\phi^2\big),
	\label{extmetric}
\end{equation}
which in the absence of electric field, reduces to the following Schwarzschild metric \citep{Misner}
\begin{equation}
	ds^2 = \Big(1-\frac{2M}{r}\Big)dt^2 - \Big(1- \frac{2M}{r}\Big)^{-1}dr^2-r^2\big(d\theta^2 + sin^2\theta d\phi^2\big).
\end{equation}
\subsection{Physical Conditions}
The study of relativistic objects in general relativity is achieved by solving the Einstein-Maxwell system of equations and imposing conditions for physical acceptability. These conditions are:
\begin{enumerate}
	\item The internal metric and the external metric must be continuous about the surface $r=R$. 
	\item The solution must not have singularities, i.e., for $0\le r\le R$ the metric functions, the charge, the density, and pressures are non-negative.
	\item The charge should be a monotonically increasing function of radius, towards the surface of the sphere.
	\item The pressures and density must be positive and monotonically decreasing, with its maximum value in the center.
	\item Radial pressure should vanish at boundary.
	\item Causality conditions: The condition of causality must not be violated, i.e., the magnitude of the speed of sound must be less than the speed of light. 
	\item Well behaved condition: The speed of sound should be a monotonic function decreasing towards the surface.
	\item Gravitational redshift, as well as surface redshift, should be positive and finite. Gravitational redshift should monotonically be decreasing in nature with the increase of radius, whereas surface redshift should increase with radius and meet the value of gravitational redshift at the surface.
	\item Energy conditions: Some conditions have been imposed on the linear relationship of energy density and pressure of anisotropic compact stars, which are collectively known as energy conditions. The solutions are required to fulfill all the energy conditions simultaneously.
	\item Equilibrium condition: The solution must satisfy the balancing force equation, known as the TOV-equation.
	\item Stability conditions: 
	\begin{enumerate}
		\item  via Cracking: The condition for potentially stable region is $-1<\delta v_s<0$, wheras $0<\delta v_s<1$ is potentially unstable region \citep{Abreu}.
		\item via Adiabatic Index: Any stable configuration should have $\Gamma\ge \frac{4}{3}+\frac{19}{21}\frac{M}{R}$ \citep{Moustakidis}, where $\Gamma$ denotes the adibatic index of the compact object.
		\item via Harrison–Zeldovich–Novikov criterion: A configuration is stable iff its mass increases with increase in central density \citep{Harrison,Zeldovich}. 
	\end{enumerate}
\end{enumerate}

\subsection{The classification scheme}
Unlike the exterior metric for a spherically symmetric charged distribution of matter, which is uniquely given by  Reissner-Nordstr\"{o}m (or Schwarzschild) metric, the interior solutions are far from unique. Interior fluid solutions have been studied by different authors. There are a fair number of physically viable, static, spherically symmetric exact solutions of Einstein’s equations which have been written down in closed-form. 

It is not easy to achieve a solution of the field equations because of its nonlinearity. We have already seen that equations of a stellar structure are coupled differential equations that have more unknowns than the number of equations. Supplementary equations or data is needed in order to solve it numerically. 
Simple approximation or assumption helps us to obtain a solution to the stellar structure problem without resorting to a computationally expensive full solution of the coupled differential equations of stellar structure. 
There could be different approaches to obtain the solution. One of them is to consider that stellar models are homologous; i.e., all physical variables in stellar interiors scale the same way with the independent variable measuring distance from the stellar center (the interior mass at some point specified by m(r)). The stellar mass (m) is used as a scaling factor. 
A second approach is to suppose that at some distance $r$ from the stellar center, the pressure and density are related by $p_r=p_r(\rho)$. 
Another approach could be to consider that both the metric potentials are related.

In this article, we will classify the interior anisotropic fluid solutions for the Reissner-Nordstr$\ddot{\text{o}}$m (or Schwarzschild) metric according to the technique used to generate these solutions. i.e., whether the components of energy-momentum tensor ($\rho$, $p_r$ and $p_t$) have been considered independent to each other or some constraints are imposed on the system of Einstein-Maxwell field equations that establish some relation between these physical parameters. 

We have three equations (\ref{fe1}-\ref{fe3}) for six unknown functions ($\lambda$, $\nu$, $q$, $\rho$, $p_r$ and $p_t$). When $\rho$, $p_{r}$ and $p_{t}$ are assumed to be independent of each other, one can freely choose three of the six unknowns and solve for the remaining ones. But the model will be physically realistic if regularity, matching, and stability conditions are satisfied too.
The relation between physical parameters can be established by assuming an equation of state, or considering the spacetime admitting conformal motion, or supposing embedding class one spacetime obeying the Karmarkar condition. The first assumption relates the pressure to the density of the star and the latter two assumptions establish relations between both the metric potentials.

In the present article, we have discussed different ways to generate models for static and spherically symmetric anisotropic stellar objects. Section \ref{sec2} deals with the solutions having mutually independent components of the energy-momentum tensor. Sections \ref{sec3}, \ref{sec4} and \ref{sec5} deals with the solutions where these components are assumed to be related. In section \ref{sec3} we have discussed the role of an equation of state (EoS). In section \ref{sec4} we have discussed the solutions based on the condition for the existence of conformal motion. In section \ref{sec5} solutions of embedding, class one spacetime have been discussed. Finally, in section \ref{sec6} we have concluded the article. 
\clearpage
\section{Solutions having mutually independent components of the energy-momentum tensor} \label{sec2}
In this section, we have studied the interior anisotropic fluid solutions for the Reissner-Nordstr$\ddot{\text{o}}$m (or Schwarzschild) metric based on a new classification scheme, considering that the components of the energy-momentum tensor are mutually independent. The presence of six unknown functions and just three essential field equations allows one to specify any three fluid characteristics and solve for the remaining. 
Even though we have $^6C_3$ possibilities, there are two main ways to generate a model. The first way is to consider $q$ as a given function and the second one is when $q$ is not given beforehand. In the first case, we are free to choose any throughout positive, monotonically increasing function which is regular at origin, as a charge function of the stellar model. This is simply the case of charged generalization of a stellar model for neutral fluid, thus rather probable that this electrified model will be physically realistic if so the uncharged one is. In the second case, we obtain the charge function using field equations by considering any three fluid characteristics other than $q$ as priori. We have no control over the charge function in this case.

Our classification scheme depends on which three of the fluid characteristics are given functions. For example, ($\lambda,\nu, q$) is the case of given metrics and charge function, and the other three fluid characteristics are found from the equations. This does not mean that solutions are distributed among groups that do not overlap. Any solution, after $\lambda$, $\nu$, and $q$ are known, maybe put into this class. 
When the components of the energy-momentum tensor are given mutually independent, we can classify the solutions as shown in Table \ref{t1} and Table \ref{t2}. The uncharged solutions can be considered as the case of known $q$ where charge function is given as a zero function.
\begin{table}[h]
	\centering
		\caption{\label{t1} Classified groups of solution with known $q$.}
		\begin{tabular}{|l|l|}
			\hline
			Cases & solution \\
			\hline
			($\lambda,\nu$); ($\rho,\nu$)	& \citep{Maurya15,Maurya18,Maurya19} \\
			\hline		
			($\lambda,p_r$); ($\rho,p_r$); ($\nu, p_r$)& ... \\
			\hline
			($\lambda,\Delta$); ($\rho,\Delta$); ($\nu,\Delta$)	&\citep{Karmakar07,amit19,maurya17}  \\
			\hline
			($\lambda,p_t$); ($\rho,p_t$); ($\nu, p_t$)& ... \\
			\hline
			($p_r,p_t$); ($p_r,\Delta$); ($p_t,\Delta$)&none \\
			\hline
		\end{tabular}

		\caption{\label{t2} Classified groups of solution when $q$ has to be obtained.}
		\begin{tabular}{|l|l|}
			\hline
			Cases & solution \\
			\hline
			($\lambda,\nu,\rho$)&...\\
			\hline
			($\lambda,\nu,p_r$); ($\lambda,\nu,p_t$); ($\lambda,\nu,\Delta$)&...\\
			\hline
			($\rho,\nu,p_r$) &...\\
			\hline
			($\rho,\nu,p_t$); ($\rho,\nu,\Delta$)&...\\
			\hline
			($\rho,p_r,p_t$); ($\rho,p_r,\Delta$); ($\rho,p_t,\Delta$)& none\\
			\hline
		\end{tabular}
\end{table}

Lets consider the transformation $Y=e^{-\lambda}$, then eqs. (\ref{fe1})-(\ref{fe4}) get transformed into following forms:
\begin{eqnarray}
	&Y'+\frac{Y}{r}=\frac{1}{r}-c^2\kappa r \rho-\frac{q^2}{r^3}& \label{fe11}\\
	\nonumber	&\nu'=\frac{r}{Y}\left[\frac{1-Y}{r^2}+\kappa p_r- \frac{q^2}{r^4}\right] \ \ \ \text{or} \ \ \ \kappa p_r=\frac{\nu'}{r}Y-\frac{1-Y}{r^2}+\frac{q^2}{r^4} \ \ \\ 
	&\text{or} \ \ \ Y=\frac{-r^2}{1+r\nu'}\left[\frac{q^2}{r^4}-\kappa p_r-\frac{1}{r^2}\right]& \label{fe21}\\
	&\kappa p_t=\frac{Y'}{2r}+\left(\frac{\nu''}{2}-\frac{\lambda' \nu'}{4}+\frac{\nu'^2}{4}+\frac{\nu'}{2r}\right)Y-\frac{q^2}{r^4} & \label{fe31}\\
	&\kappa \Delta=-\left(\frac{\nu'}{4}+\frac{1}{2r}\right)Y'+\left(\frac{\nu''}{2}+\frac{\nu'^2}{4}-\frac{\nu'}{2r}-\frac{1}{r^2}\right)Y+\frac{1}{r^2}-\frac{2q^2}{r^4}& \label{fe41}
\end{eqnarray}

Above equations shows that \ref{fe1}-\ref{fe4} are linear differential equations with respect to $e^{-\lambda}=Y$ and equation (\ref{fe3}) is a  linear differential equation with respect to $\nu$. Thus, it will not be difficult in general to solve the equations for these parameters when other parameters corresponding to the equations are given. Now, consider the case of given $q$. In this case, equation (\ref{fe1}) gives the expression for energy density whenever $\lambda$ is known and vice-versa. Eq. (\ref{fe21}) shows that eq. (\ref{fe2}) is the simplest generating function of any of $\nu$, $\lambda$ and $p_r$, provided that the other two are known.

Using transformation $Z=e^{\nu/2}$, eqs. (\ref{fe31}) and (\ref{fe41}) can be transformed into a linear second order homogeneous differential equation as follows:
\begin{eqnarray}
	&YZ''+\frac{rY'+2Y}{2r}Z'+\left[\frac{Y'}{2r}-\kappa p_t-\frac{q^2}{r^4}\right]Z=0 \label{fe311}\\
	&YZ''-\frac{rY'+2Y}{2r}Z'+\left[-\frac{Y'}{2r}+\frac{1-Y}{r^2}-\kappa \Delta-\frac{2q^2}{r^4}\right]Z=0&\label{fe411}
\end{eqnarray}
The above two equations are solvable for suitable choices of $Y$ and $p_t$ or $\Delta$. Thus, with the help of eqs. (\ref{fe31}) - (\ref{fe411}) wecan say that, like $p_r$, the expressions (\ref{fe3}) for $p_t$ and (\ref{fe4}) for $\Delta$ may be considered as a generating functions for stellar models, when two of the quantities $\lambda$, $\nu$ and $p_t$ or $\Delta$ are known. 
When $p_r$ and $p_t$ are known, one can use eqs. (\ref{fe2}) and (\ref{fe3}) collectively to obtain the other fluid characteristic. 

($\lambda,\nu, q$) is comparatively the simplest case of generating solutions. As $\lambda$, $\nu$, and $q$ are given functions and there is control over them; they can be chosen as regular, positive, and comparatively simple. However, the other three functions are usually more complex and are not always physically realistic. One completely lost control over pressures and density in this case and has to proceed further by trial and error.

We can categorize the solutions as per Table \ref{t2} in a similar manner when $q$ is not given priori. However, in this case, the second approach, i.e., imposing constraints on the system of field equations is more popular. One can proceed with the first approach too.\\\\
Tables \ref{t3}-\ref{t5} give the name and reference of the solution studied, along with the corresponding metric. We have named the solutions with the authors name (with a number to differentiate between multiple publications by one author), followed by a solution number. 
We do not claim that this table is complete.

\begin{sidewaystable}[h]
	\centering
	\caption{\label{t3} ($\lambda,\nu,q$); ($\rho,\nu,q$)}
	\tiny	\begin{tabular}{|l|l|}
		\hline
		name [ref.] & metric  \\ 
		\hline
		M-G-R-D	&$ds^2=e^{(1-\alpha+Cr^2)^4}dt^2-\frac{7+14Cr^2+7C^2r^4}{7-10Cr^2-C^2r^4}dr^-r^2d\Omega^2$;  \\
		\citep{Maurya15}	&$\alpha>0, 0<Cr^2<\frac{\sqrt{\alpha^2+8\alpha+272}-16+K}{4}$\\
		\hline
		M-B-H \citep{Maurya18}&$ds^2=(1+Cr^2-\beta Cr^2)dt^2-\frac{1+2Cr^2}{1+Cr^2}dr^2-r^2d\Omega^2$\\
		\hline
		M-B-J-K-P-P Ia&$ds^2=-{f^2(r)}dt^2+\frac{K+Cr^2}{K(1+Cr^2)}dr^2+r^2d\Omega^2$; $K<0$\\
		\citep{Maurya19}&$f(r)=\frac{1}{1+n^2}\{\cosh{(nx)}(A\sin x+Bn\cos x)+\sinh{(nx)}(B\sin x+An\cos x)\}$; $x=\sin^{-1}\sqrt{\frac{K+Cr^2}{K-1}}$\\
		\hline
		M-B-J-K-P-P Ib	&$ds^2=-{f^2(r)}dt^2+\frac{K+Cr^2}{K(1+Cr^2)}dr^2+r^2d\Omega^2$; $K<0$\\
		\citep{Maurya19}&$f(r)=\frac{1}{1-n^2}\{\sin{x}(B\sin{(nx)}+A\cos{(nx)})-n\cos{x}(A\sin{(nx)}-B\cos{(nx)})\}$; $x=\sin^{-1}\sqrt{\frac{K+Cr^2}{K-1}}$\\
		\hline	
		M-B-J-K-P-P Ic	\citep{Maurya19}&$ds^2=-{\left[\frac{1}{4}\{A(2x+\sin{(2x)})-B\cos{(2x)}\}\right]}^2dt^2+\frac{K+Cr^2}{K(1+Cr^2)}dr^2+r^2d\Omega^2$; $K<0$; $x=\sin^{-1}\sqrt{\frac{K+Cr^2}{K-1}}$\\
		\hline	
		M-B-J-K-P-P	Id \citep{Maurya19}&$ds^2=-{\left[A\cos x+(Ax+B)\sin x\right]}^2dt^2+\frac{K+Cr^2}{K(1+Cr^2)}dr^2+r^2d\Omega^2$; $K<0$; $x=\sin^{-1}\sqrt{\frac{K+Cr^2}{K-1}}$\\
		\hline	
		M-B-J-K-P-P IIa	&$ds^2=-{f^2(r)}dt^2+\frac{K+Cr^2}{K(1+Cr^2)}dr^2+r^2d\Omega^2$; $K<0$\\
		\citep{Maurya19}&$f(r)=\frac{1}{1+n^2}\{\cosh{x}(B\sin{(nx)}+A\cos{(nx)})+n\sinh{x}(A\sin (nx)-B\cos(nx))\}$; $x=\cosh^{-1}\sqrt{\frac{K+Cr^2}{K-1}}$\\
		\hline
		M-B-J-K-P-P IIb	&$ds^2=-{f^2(r)}dt^2+\frac{K+Cr^2}{K(1+Cr^2)}dr^2+r^2d\Omega^2$; $K<0$\\
		\citep{Maurya19}&$f(r)=\frac{1}{1-n^2}\{\cosh{x}(B\sinh{(nx)}+A\cosh{(nx)})-n\sinh{x}(A\sin (nx)-B\cos(nx))\}$; $x=\cosh^{-1}\sqrt{\frac{K+Cr^2}{K-1}}$\\
		\hline
		M-B-J-K-P-P IIc \citep{Maurya19}&$ds^2=-{\left[\frac{1}{4}\{B(-2x+\sinh{(2x)})+A\cosh{(2x)}\}\right]}^2dt^2+\frac{K+Cr^2}{K(1+Cr^2)}dr^2+r^2d\Omega^2$; $K<0$; $x=\cosh^{-1}\sqrt{\frac{K+Cr^2}{K-1}}$\\
		\hline	
		M-B-J-K-P-P IId \citep{Maurya19}&$ds^2=-{\left[-B\sin x+(Ax+B)\cosh x\right]}^2dt^2+\frac{K+Cr^2}{K(1+Cr^2)}dr^2+r^2d\Omega^2$; $K<0$; $x=\cosh^{-1}\sqrt{\frac{K+Cr^2}{K-1}}$\\
		\hline
		K-B \citep{Bharti}&$A^2\sqrt{\mid1-Y^2\mid}frac{(\alpha+\beta Y)^2}{Y^2} \bigg [\frac{\alpha}{\beta^3}f(Y)+\frac{B}{A}\bigg]^2 dt^2+\frac{K+Cr^2}{K(1+Cr^2)}dr^2+r^2d\Omega^2$; $K<0$ or $K>1$; $Y=\sqrt{\frac{K+x}{K-1}}$;\\
		& $f(Y)=\frac{\sec^2t-\cos^2t}{2}+\log{\cos^2t}$ with $t=\tan^{-1}\sqrt{\frac{\beta Y}{\alpha}}$.\\	
		\hline
	\end{tabular}
\end{sidewaystable}
\begin{sidewaystable}[h]
	\centering
	\caption{\label{t4} ($\lambda,\Delta,q$);($\rho,\Delta,q$) }
	\begin{tabular}{|l|l|}
		\hline
		name [ref.] & metric  \\
		\hline
		K-M-S-M \citep{Karmakar07}	&$ds^2=A\left[\frac{cos(\sqrt{\delta(1-\alpha)+2}+1)cos^{-1}\sqrt{\frac{\lambda}{\lambda-1}}+\delta}{\beta+1}-\frac{cos(\sqrt{\delta(1-\alpha)+2}-1)cos^{-1}\sqrt{\frac{\lambda}{\lambda-1}}+\delta}{\beta-1}\right]^2dt^2-\frac{R^2+\lambda r^2}{R^2-r^2}dr^2-r^2d\Omega^2$\\
		\hline
		M-M-K-P \citep{amit19}	&$ds^2=\sqrt{\frac{4(1+Cr^2)}{3}},$$(f(r)+1)^2\left[A\lvert\frac{f(r)+1}{f(r)-1}\rvert^{\alpha-1/2}+B\lvert\frac{f(r)+1}{f(r)-1}\rvert^{-\alpha-1/2}\right]dt^2-\frac{7(K+Cr^2)}{7+4Cr^2}dr^2-r^2d \Omega^2$, $C>0$,\\
		&$f(r)=\sqrt{\frac{7+4Cr^2}{3}}$, $\alpha=\frac{\sqrt{5\beta+9}}{4}$.\\
		\hline
	\end{tabular}

	\centering
	\caption{\label{t5} ($\nu,\Delta,q$)}
	\begin{tabular}{|l|l|}
		\hline
		name [ref.] & metric  \\
		\hline
		M-G-R\citep{maurya17}&$ds^2=e^{-\log{\frac{1-Cr^2}{B}}}dt^2-e^{1-f_1(r)+f_2(r)}dr^2-r^2d\Omega^2$\\
		&$f_1(m)=1-K(1-Cr^2)^2e^{2Cr^2}[\frac{(Cr^2)^{N+2}}{2}(a-bCr^2)^m+Cr^2(1-Cr^2)\Pi_{a,b,C,m,i,N}(r)]+2(\beta+3)Cr^2$ \\
		&$(1-Cr^2)^3e^{2(Cr^2-1)}[\phi_{C,j}(r)+\psi_{C,j}(r)]-A(1-Cr^2)^3Cr^2e^{2Cr^2}-[1+(\beta-2)Cr^2](1-Cr^2)^2$,\\
		&$f_2(r)=\frac{K(Cr^2)^{2+N}}{2}(1-Cr^2)^2(a-bCr^2)^m$, $\Pi_{a,b,C,m,i,N}(r)=\sum_{i=0}^{m}(-b)^ia^{m-i} {{m}\choose{i}} \frac{(Cr^2)^{N+i+1}}{N+i+1}$,\\
		&$\phi_{C,j}(r)=\log{(2-2Cr^2)}$, $\psi_{C,j}=E_i{(2-2Cr^2)}-\log{(2-2Cr^2)}$.\\
		\hline
	\end{tabular}
\end{sidewaystable}
\clearpage
\section{Solutions when an equation of state is assumed}\label{sec3}
In the previous section, we have discussed the case when components of the energy-momentum tensor are mutually independent. In this section, we have considered the solutions having mutually related radial pressure and density. 

To obtain a stellar model, one can assume an equation of state (EoS) and solve the field equations to explore the rest of the physical properties. An EoS intimately connects the radial pressure of the stellar fluid to its density via $p = p(\rho)$. In the study of compact objects, EoS remains a key aspect. A fruitful and physically plausible approach to stellar modeling, within the framework of classical general relativity, is to adopt an EoS a priori. When one introduces an EoS, the problem reduces to four equations to solve for six unknowns. So, one can proceed further by considering any two of the fluid characteristics as priori or by imposing some additional constraints.

Models with an equation of state are desirable in describing viable astrophysical objects. However, most of the present solutions of the Einstein-Maxwell system do not satisfy this property. There is still a debate over the structure of a star. Because of a lack of knowledge of nuclear interactions beyond the density $10^{14}g/cc$, there is no well-known EoS that describes the exact structure of a compact star.  It is not well known that the composition of a star is in terms of nuclear matter or quark matter, or it is a hybrid mixture of both distributions. Usually, the stellar core consists of softer quark matter, whereas the outer regions of a stellar structure consist of stiffer nuclear matter. Thus, it is difficult to find a single EoS for a matter distribution matching the core to the outer regions \citep{Cottam, Ozel, Rodrigues}.  

To construct physically motivated stellar models an equation of state has been considered in numerous works. Most of the earlier works were centered on imposing a linear EoS. Investigations in fundamental particle physics led to the MIT-bag model, which depends on an equation of state similar to the linear one. The bag model equation of state has been the basis for the study of most of the static relativistic models of strange stars. More sophisticated investigations of quark-gluon interactions clarified that bag model EoS represents a limiting case of more general equations of state. For example, MIT bag models with massive strange quarks and lowest order quantum chromodynamics interactions lead to some correction terms in the equation of state of quark matter. Models incorporating restoration of chiral quark masses at high densities and giving stable strange matter can no longer be accurately described by bag model EoS \citep{Mak2002}. The linear equation of state is further generalized to the quadratic EoS.
Buchdahl \citep{Buchdahl} was one of those who successfully obtained a generalization of the Newtonian polytrope.  Later Herrera and Barreto \citep{Herrera} presented a general formalism to generate relativistic polytropes with anisotropic pressure in Schwarzschild coordinates. These findings motivated further investigations on the origins of anisotropy and stability of a relativistic stellar models.
The proposal of treating dark energy and dark matter as different manifestations of a single entity leads to the Chaplygin gas model. In this model, the EoS derives from string theory. The Chaplygin EoS has subsequently modified to a more generalized Chaplygin gas EoS, and this has been employed to model dark stars in several papers. 

We have characterized the solutions of this class based on the choice of EoS. This classification scheme has been summarized in below Table \ref{teos}.
\begin{sidewaystable}[h]
	\centering
	\caption{\label{teos} Classified group of solutions generated by assuming an EoS.}
\tiny	\begin{tabular}{|l|l|}
		\hline
		Cases & Solutions \\
		\hline
		Linear EoS&\citep{Maha2006,Sharma07,Esculpi10,Takisa13,Matondo16,Takisa16,Malaver18}  \\
		\hline
		MIT EoS	& \citep{bis19,rag19,shar20,KomaMit,Goswami,Jasim21}\\
		\hline
		Quardratic EoS& \citep{bhar16,Malaver17,Sunzu18,takisa19,Malaver20,pant20,thiru21} \\
		\hline
		CFL EoS	& \citep{Rocha19,Rocha20,thiru20,singh21} \\
		\hline
		Polytropic EoS&\citep{Pandey,Thirukkanesh12,ngu15a,Ngu17,Isayev,Azam3,Noureen,Mardan}\\
		\hline
		Chaplygin EoS& \citep{rah10,bhar15,bhar18,Ortiz0,amit,Mala}\\
		\hline
		Van der Waals EoS& \citep{Mala13a,Mala13b,sunzu18,Malaver19}\\
		\hline
	\end{tabular}
\end{sidewaystable}
\clearpage
\subsection{Linear equation of state}
In the past, the simplest EoS relate the fluid density and pressure via a linear relation. An interesting study of compact objects in classical general relativity, using the Vaidya-Tikekar ansatz \citep{Vaidya}, revealed a linear relationship between the energy density and the isotropic pressure. Frieman and Olinto \citep{Frieman} and Haensel and Zdunik \citep{Haensel} already mentioned the approximation of the EOS by a linear function in 1989. Models in which quark interaction was described by an interquark potential originating from gluon exchange and by a density-dependent scalar potential which restores the chiral symmetry at high densities, the equation of state $p =p(\rho)$ can be well approximated by a linear function in the energy density $\rho$ \citep{GondekRosinska}. Linear equations of state are relevant in the modeling of static spherically symmetric anisotropic quark matter distributions \citep{Mak2002}.

In the linear equation of state, radial pressure is assumed to be some linear function of $\rho$ as
\begin{equation}
	\nonumber	p_r = \alpha \rho-\beta
\end{equation}
where $\alpha$ and $\beta$ are real constants. 

Recently, there are made several attempts to obtain anisotropic models with a linear equation of state. These include the work of Sharma and Maharaj \citep{Sharma07}, which describes the model for a class of relativistic stars,  the investigations of Esculpi and Aloma \citep{Esculpi10}, on the conformal anisotropic relativistic charged fluid spheres, models proposed by Takisa and Maharaj \citep{Takisa16}, the treatments of Matondo and Maharaj \citep{Matondo16}, charged models developed by Sunzu and Danford \citep{Sunzu17}, generalized compact star models by Malaver \citep{Malaver18}, and models of Sunzu et al. \citep{Sunzu19}, for charged anisotropic stars in general relativity. Other linear models are those found by Harko and Cheng \citep{Harko}, Sotani et. al. \citep{Sotani}, Maharaj and Chaisi \citep{Maha2006}, the study of Thirukkanesh and Maharaj \citep{Thirukkanesh08} on a charged anisotropic matter, and the investigations by Takisa and Maharaj \citep{Takisa13} and Islam et al. \citep{islam1,islam} on compact models with regular charge distribution.

In Table \ref{tl} we have given the metric used by several researchers to model an anisotropic stellar structure by employing linear EoS. We have also discussed the process used by them.

\begin{sidewaystable}[h]
	\centering
	\caption{\label{tl} Solutions obtained by considering linear EoS.}
	\tiny	\begin{tabular}{|l|l|}
		\hline
		name [ref.] & metric  \\
		\hline
		M-C&$ds^2=Br^{a/(1-a)}e^{f(r)}dt^2-\frac{1}{1-a-\frac{br^2}{3}-\frac{cr^4}{5}}dr^2-r^2d\Omega^2$, where,\\
		\citep{Maha2006}&$f(r)=\frac{f_1}{1-a}-\frac{5a\eta}{c(n^2-m^2)}\log{r}-\frac{5\eta}{4mc(n+m)}(a+(n+m)b+(n+m)^2c)\log{(r^2-n-m)}+\frac{5\eta}{4mc(n-m)}(a+(n-m)b+$\\
		&$(n-m)^2c)\log{(r^2-n+m)}$, with $n=-\frac{5b}{6c}$, $m=\frac{5}{2c}\sqrt{\frac{b^2}{9}+\frac{4c}{5}(1-a)}$, $f_1(r)=-\frac{1}{2}\log{(1-a-\frac{br^2}{3})} \ (\text{for} \ c=0) \ \& $\\
		& $\ -\frac{1}{4}\log{(1-a+\frac{5b^2}{36c}-(r^2+\frac{5b}{6c})^2)}+ \frac{b\sqrt{\frac{5}{c}}}{12\sqrt{1-a+\frac{5b^2}{36c}}}\tanh^{-1}{\Big( \frac{r^2b\sqrt{\frac{5}{c}}}{6\sqrt{1-a+\frac{5b^2}{36c}}}\Big)} \ (\text{for} \ c\ne0)$ and $0\le \eta \le1$.\\
		\hline
		S-M&$ds^2=-B(1+ar^2)^\alpha[1+(a-b)r^2]^ne^{\frac{-acr^2}{2(a-b)}}dt^2+\frac{1+ar^2}{[1+(a-b)r^2]}dr^2+r^2d\Omega^2$\\
		\citep{Sharma07}&where, $n=\frac{(5ab-2a^2-3b^2)\alpha+(a-b-c)b}{2(a-b)^2}$\\
		\hline
		T-M I&$ds^2=-A^2D^2(1+aCr^2)^{\frac{2a\alpha-K(1+\alpha)}{2a}}e^{2f(r)}dt^2+(1+aCr^2)dr^2+r^2d\Omega^2$\\
		\citep{Thirukkanesh08}&with, $f(r)=\frac{r^2}{8}[-KC(1+\alpha)-2\beta+aC(2+2\alpha-\beta r^2)]$\\
		\hline
		T-M II&$ds^2=-A^2D^2(1+aCr^2)^{2m}[1+(a-b)Cr^2]^{2n}e^{\frac{-a\beta Cr^2}{2C(a-b)}}dt^2+\frac{1+aCr^2}{1+(a-b)Cr^2}dr^2+r^2d\Omega^2$\\
		\citep{Thirukkanesh08}&with, $m=\frac{2b\alpha-K(1+\alpha)}{4b}$ \ $\&$ $n=\frac{1}{8bC(a-b)^2}[2a^2C(K(1+\alpha)-2\alpha \beta)-abC(5K(1+\alpha)-2b(1+5\alpha))+b^2(3KC(1+\alpha)-$\\
		&$2bC(1+3\alpha)+2\beta)]$\\
		\hline
		E-A I \citep{Esculpi10}&$ds^2=ar^2dt^2-[\frac{1+\alpha}{2(1+2\alpha-C)}+\frac{8\pi \beta (1+Cr^2)}{6(\alpha+1)}]dr^2-r^2d\Omega^2$\\
		\hline
		E-A II \citep{Esculpi10}&$ds^2=ar^2dt^2-[\frac{1+\alpha}{3\alpha+1}+\frac{8\pi \beta r^2}{3(\alpha+1)}-\frac{(\alpha+1)Q_0}{3\alpha+2n-1}(\frac{r}{R})^{2n}]dr^2-r^2d\Omega^2$\\
		\hline
	\end{tabular}
\end{sidewaystable}

\textbf{Working Rule:}\\
\textbf{Maharaj $\&$ Chaisi} \citep{Maha2006}: They have considered the energy density as $\rho = \frac{a}{r^2} +b+r^2$, where a, b and c are constants. For this form of energy density and the linear equation of state with $\beta=0$, they have integrated the Einstein field equations and exhibited an exact solution following the integration procedure. But, this uncharged model with a linear EoS has the unrealistic feature of vanishing energy density at the boundary. \\
\textbf{Esculpi $\&$ Aloma} \citep{Esculpi10}: Considering an electrically charged and anisotropic spherically symmetric distribution of matter, the authors have obtained two new families of compact solutions. They also considered that the static inner region of the model admits a one-parameter group of conformal motions. To obtain the first family of solutions, besides assuming a linear equation of state to hold for the fluid, they assumed the ratio of tangential pressure to radial pressure is a constant factor $C$, which measures the grade of anisotropy. In this case, they obtained the charge distribution within the sphere from the field equation of state. 
The second family of solutions is obtained by assuming the electrical charge inside the sphere as a known function of the radial coordinate ($Q=Q_0\frac{r^n}{R^n}$). 
They claimed that the obtained solutions could model some stage of the evolution of strange quark matter fluid stars.\\
\textbf{Thirukkanesh $\&$ Maharaj} \citep{Thirukkanesh08}: In this paper, the authors have found solutions for the Einstein-Maxwell system of equations with a linear equation of state for charged anisotropic matter distributions. They made specific choices for the gravitational potential $\lambda$ and electric-field intensity $E$ as, $e^\lambda=\sqrt{\frac{1+aCr^2}{1+(a-b)Cr^2}}$ and $E=\frac{\sqrt{KC(3+aCr^2)}}{1+aCr^2}$, where $a$, $b$, $K$ and $C$ are real constants. To avoid negative energy densities they have taken $b\ne0$. They have obtained models considering the cases when $a=b$ (T-M I) as well as when $a \ne b$ (T-M II). For $K=0$ this model reduces to a solution for uncharged anisotropic stars. Furthermore, they have shown that the obtained models contain the uncharged anisotropy models of Sharma and Maharaj \citep{Sharma07} and Lobo \citep{Lobo} which describe quark stars and strange matter stars. On setting $\beta=\alpha \rho_R$ and $C=1$, where $\rho_R$ is the density at the surface $r=R$, the model reduces to Sharma and Maharaj model and by setting $\beta=0$, $a=2b$ and $C=1$, Lobo model can be obtained. Also, for particular parameter values, their metric reduces to metrics corresponding to the isotropic uncharged Einstein model.\\
\textbf{Sharma $\&$ Maharaj} \citep{Sharma07}: The authors have constructed a new model for compact stars where pressure anisotropy is present and the EOS is linear. Unlike previous models, they didn't choose the form of the anisotropic parameter a priori in this model. Their analysis depends critically on their particular choice of the mass function, using which they obtained one of the metric functions as $e^\lambda=\frac{1+ar^2}{1+(a-b)r^2}$. Along with this choice they have considered the parameter $\beta$ in linear EoS as, $\beta=\alpha \rho_R$, where $0\le \alpha \le 1$ is a constant which is related to the sound speed $\frac{dp_r}{d\rho} =\alpha$, and $\rho_R$ is the density at the surface $r = R$. The obtained solutions are found non-singular and well-behaved in the stellar interior. 
\clearpage
\subsection{MIT Bag Model EoS}
Current observational data indicate the possibility of existence of compact stars composed of deconfined quarks known as strange stars. Nowadays researcher has mentioned about the possible formation of quark stars (QS) and hybrid star, where quark matters are present at the interior. The simplest EoS that governs these quark matters is the MIT Bag Model, where quark matters are considered as non-interacting Fermi gas bound inside a hard-sphere of nucleons. This EoS describing quark matters is also relatively softer. 
Since we do not observe free quarks, in an attempt to describe the quark confinement mechanism, the phenomenological MIT Bag model is proposed, where one assumes that the quark confinement is due to a universal pressure known as the ‘bag pressure’ at the boundary of the region containing quarks. The first description of the bag model appeared in 1967 with the publication of the work by P. N. Bogoliubov. Even though this model has few defects, the major one being the violation of the energy-momentum conservation, it was surprising that it estimate the mass of the nucleus with one of its quarks very close to the experimental value. With the passing years, the paper was almost forgotten. Seven years later, in 1974, a group of five scientists from MIT reinvented the Bag Model. This enhanced version of Bogoliubov's work, referred to as the MIT bag model, doesn't violate the energy-momentum conservation. Even now, 46 years later, the MIT bag model is used to calculate the properties of the nucleus.

The MIT Bag model motivated by work in quantum chromodynamics (QCD) centered on quark confinement and asymptotic freedom \citep{Benve89} is essentially a linear EoS of the form \begin{equation}
	p_r=\frac{(\rho-4B)}{3}
\end{equation}
where $B$ is the Bag constant \citep{rah12,bis19}. This EoS cannot adequately account for the deconfinement of quarks at high-density \citep{dey98}. Furthermore, the magnitude of the Bag constant is not absolute but rather depends on the compact objects being modeled \citep{kal13}. For stability, one assumes the bag constant within a particular range, even though density and temperature-dependent bag models have been proposed where the bag constant can take a much wider range of values. An ultra-dense strange (quark) star with the MIT Bag model with Mak and Harko \citep{Mak2002} density profile was studied by Deb et al. \citep{Deb}. There is various literature available regarding the MIT Bag model EOS \citep{Brilenkov, Panda, Bhar,bibi}. The MIT Bag model has been employed in several recent works to generate models of anisotropic spheres \citep{bis19,rag19,shar20,KomaMit,Goswami,Jasim21}.

Here, in Table \ref{tm} we have given some metrics used in different anisotropic stellar model, employing MIT bag model EoS. We have also discussed these solutions briefly.  
\begin{sidewaystable}[h]
	\centering
	\caption{\label{tm} Solutions obtained by considering MIT bag model EoS.}
	\begin{tabular}{|l|l|}
		\hline
		name [ref.] & metric  \\
		\hline
		K-S I&$ds^2=-A^2(a+\sqrt{C}(b+1)r)^{2}dt^2+\frac{3(2a+3\sqrt{C}(b+1)r)}{3(2a+\sqrt{C}(b+1)r)-Br^2(4a+3\sqrt{C}(b+1)r)+3\alpha {(\sqrt{C}r)}^3}dr^2+r^2d\Omega^2$\\
		\citep{KomaMit}&\\
		\hline
		K-S II&$ds^2=-A^2(a+1+bCr^2)^4dt^2+\frac{315(a+1+bCr^2)^2(a+1+3bCr^2)}{9[35(a+1)^2(a+1+bCr^2)+b^2C^2r^4(21(a+1)+5bCr^2)-f(r)]}dr^2+r^2d\Omega^2$\\
		\citep{KomaMit}&where, $f(r)=2r^2[21(a+1)^2(5a+5+9bCr^2)+5b^2C^2r^4(27+27a+7bCr^2)]+2\alpha C^2r^4$\\
		&$[63(a+1)^2+90(a+1)bCr^2+35b^2C^2r^4]$\\
		\hline
		G-S-C&$ds^2=-A^2[\frac{cos[(\alpha+1)\cos^{-1}(\theta)+B]}{\alpha+1}-\frac{cos[(\alpha-1)\cos^{-1}(\theta)+B]}{\alpha-1}]^2dt^2+\frac{R^2+a^2r^2}{R^2-r^2}dr^2+r^2d\Omega^2$\\
		\citep{Goswami}&with, $\theta=\sqrt{\frac{aR^2}{(a+1)(R^2-r^2)}}$, $\alpha=\sqrt{n+(n-1)C}$\\
		\hline
	\end{tabular}
\end{sidewaystable}

\textbf{Working Rule:}\\
\textbf{Komathiraj $\&$ Sharma} \citep{KomaMit}: In this paper, they have presented two new classes of solutions for a compact star, interior strange matter composition of which admits the MIT Bag model EoS.  To obtain the solutions for the Einstein-Maxwell system for a spherical object composed of quark matter in the presence of local anisotropy, they have considered a Durgapal and Bannerji form for one of the metric variables as $e^{\nu}=A(a+b(Cr^2)^{1-m}+(Cr^2)^m)^{n}$, where $A$, $C$, $a$, $b$, $m$ and $n$ are constants.  Also, they assumed the radial fall-off behavior of the anisotropic parameter and considered $\Delta=\frac{2\alpha C}{b+a(Cr^2)^{m-1}+(Cr^2)^{2m-1}}$, where $\alpha$ is again an arbitrary constant. By considering the cases: (i). $m=\frac{1}{2}$, $n=1$ and (ii). $m=0$, $n=2$, they obtained their different class of solutions.\\ 
\textbf{Goswami et. al.} \citep{Goswami}: In this paper, the authors have studied anisotropic strange star with the MIT bag model EoS by considering Vaidya-Tikekar \citep{Vaidya} metric $e^{\lambda}=\sqrt{\frac{R^2+a^2r^2}{R^2-r^2}}$ as an ansatz to prescribe a suitable spheroidal geometry of the physical 3-space of the stellar configuration, where $a$ is the spheroidicity parameter and $R$ is a geometrical parameter. They used the value of Bag parameter ($B$) within the range of $57.55 \ \text{MeV/fm}^3 \le B \le 95.11 \ \text{MeV/fm}^3$ for stable strange matter relative to neutron. 
\clearpage
\subsection{Quardratic equation of state}
Quadratic equation of state is a simple generalization of the linear relation between the energy density and radial pressure.  It is softer at low densities and stiffer at high densities, and may be appropriate for describing a hybrid star. It offers the possibility for augmenting the pressures and densities within the core of neutron stars of the order greater than or equal to $2M_{\odot}$. It permits a more general behavior in the matter distribution and greater complexity in the model. The reason for this increase in complexity of the model is the non-linearity of the radial pressure in terms of energy density. Linear EoS like the MIT bag model EoS can only model quark stars, whereas, with quadratic EoS, we can model both quark and neutron stars. Hence the solutions obtained using quadratic EoS has more advantages in modeling compact stars. It can be reduced to a linear equation of state like the MIT bag model.
To solve the field equations via quardratic EoS, it is assumed that the radial pressure $p_r$ and the matter density $\rho$ is related by a quadratic equation of state given by
\begin{equation}
	\nonumber	p_r = \alpha \rho^2 + \beta \rho - \gamma
\end{equation}
where, $\alpha$, $\beta$ and $\gamma$ are positive constants. The above equation represents the first terms of the Taylor expansion of any equation of state of the form $P = P(\rho)$ about $\rho=0$.

Ananda and Bruni’s \citep{Ananda} work in the relativistic dynamic study of Robertson Walker models using a non-linear quadratic EoS opened a new door of research. They claimed that at the singularity found in the brane scenario, the behavior of the anisotropy can be recreated in the context of general relativity, by considering a quadratic term in the EoS. The Einstein-Maxwell field equations were brought under the application of quadratic EoS.  Capozzelio \citep{Capozziello} pointed out that the observational constraint impact along with quadratic EoS on dark energy. The work of Nojiri et al. \citep{Nojiri} declared that quadratic EoS can become a nice tool in describing dark energy or unified dark matter in a more generous way.  Rahaman et al. \citep{Rahman} showed how an electron can be modeled precisely instead of spherically symmetric charged perfect fluid distribution of matter characterized by quadratic EoS. Following this work of Rahaman et al. \citep{Rahman}, Firoze and Siddiqui \citep{Feroze} introduced “Charged anisotropic matter” into quadratic EoS and obtained a new class of static spherically symmetric models of the relativistic star.  Maharaj and Takisa \citep{Maharaj} found a new exact solution of the Einstein-Maxwell equation, for charged anisotropic matter distribution, in terms of quadratic EoS. Unification of vacuum, dark energy, and radiation \citep{ChavanisA} was also brought under active consideration through a cosmological model based on quadratic EoS. Later Chavanis \citep{ChavanisA,ChavanisB} considered early inflation, intermediate decelerating expansion, late accelerating expansion as quadratic EoS and developed a model. Focusing on the interior of a static and spherically symmetric compact anisotropic star, Sharma and Ratanpal  \citep{shar13} presented a model admitting EoS in quadratic form. After Maharaj and Tasika \citep{Maharaj}, Malaver \citep{Malaver} solved the Einstein-Maxwell system of equations for anisotropic compact objects using the quadratic EoS. In 2015 Adhav et. al. \citep{adhav} has considered a quadratic EoS, including its importance in the braneworld models and the study of dark energy and general relativistic dynamics for different models, to present a symmetric cosmological model. In their article Bhar et al. \citep{bhar16} have assumed a quadratic EoS and solved Einstein’s field equations for the first time in Tolman VII \citep{Tolman39} background. Later Malaver \citep{Malaver17} has obtained new exact solutions to the Maxwell-Einstein system for charged anisotropic matter with the quadratic EoS in static spherically symmetric spacetime using Tolman VII form, for the gravitational potential in presence of an electromagnetic field. Most of the discussed models generated with a quadratic EoS are charged and have non-vanishing anisotropy. Sunzu and Thomas \citep{Sunzu18} established neutral models with a quadratic EoS, which has vanishing anisotropy. 
For very recent models of anisotropic models of compact objects using a quadratic EoS, one can see \citep{takisa19,Malaver20,pant20,thiru21}. Pant et al \citep{pant20} used a novel approach and constructed core-envelope models for compact objects with the core obeying a linear EoS and the outer matter regions of the star described by a quadratic EoS. They checked the stabilty of such composite configuration and  proved that it meets all the physical requirements for a realistic stellar structure. Thirukkanesh et al. \citep{thiru21} used the Vaidya–Tikekar gravitational potential together with the quadratic equation of state to provide a favorable model. They have shown that in comparison with other EoS, the quadratic equation of state mimics the colour-flavor-locked EoS more closely than the linear equation of state. Recently, Pant and Fuloria \citep{fuloria} presented a comparative analysis of dense stellar models governed by linear and quadratic equations of state.

From the above discussion, we can conclude that quadratic EoS plays an important and interesting role in many areas including dark energy and dynamics of different models in general relativity. In Table \ref{tq} we have given some metrics which have been used to generate models for anisotropic stellar structure by a quadratic EoS. A brief discussion of these solutions is also given below.
\begin{sidewaystable}[h] 
	\centering
	\caption{\label{tq} Solutions obtained by considering quardratic EoS.}
	\tiny	\begin{tabular}{|l|l|}
		\hline
		name [ref.] & metric  \\
		\hline
		N-M-R&$-K^2(a-br^2)^{2n_1}(a+br^2)^{2(n_2-1)}e^{2f(r)}dt^2+\frac{1}{(a+br^2)^2}(dr^2+r^2d\Omega^2)$\\
		\citep{ngu15b}&where, $n_1=\frac{1}{256\pi ab^2}\{(1+\beta)-3\eta[b^2c^2+ad(ad-48b^3)]-b^2d\alpha(3C+dr^2)r^2\}$,\\
		&$n_2=\frac{1}{256\pi ab^2}\{16\pi ab^2(1+\beta)\{16\pi ab^2(1+\alpha)[b(24b^2+C)-ad]-256\pi^2b^4\gamma+a^2\eta[a^2d^2-b(24b^2+C)(2ad-bC-24b^3)]\}$\\
		&$\&$ $f(r)=\frac{r^2}{384\pi b^4} \{48\pi b^2 d(1+\beta)-3\eta[b^2c^2+ad(ad-48b^3)]-b^2d\alpha(3C+dr^2)r^2\}$\\
		\hline
		B-S-P   & $ds^2=-f(r)dt^2+(1+ar^2)^2dr^2+r^2 d\Omega^2$, where, $f(r)=\Big[\frac{(2+ar^2)^2}{4}+\frac{\alpha}{2 \kappa}f_1+\frac{\beta}{4}f_2 -\frac{\kappa \gamma}{6a}(1+aCr^2)^3+b\Big]$, with,\\
		\citep{bhar16} & $f_1=[a^2  r^2-\frac{a}{3(1+aCr^2)^3}\{55+66ar^2+27a^2r^4\}2A\log(1+ar^2)]$ \ $\&$ $f_2=[1+ar^2(4+ar^2)+8\log{1+ar^2}]$.\\
		\hline
		Malaver & $ds^2=-A^2k^2(1-aCr^2+bC^2r^4)^{2f(r)}e^{2g(r)}dt^2+\frac{1}{(1-aCr^2+bC^2r^4)}dr^2+r^2d\Omega^2$ \\
		\citep{Malaver17}&where, $f(r)=\frac{10\alpha f_2 -5\beta-5\alpha aC-1}{8}$, $g(r)=\frac{\sqrt{a^2-4b}h(r)+n\tan^{-1}{\Big(\frac{2bCr^2-a}{\sqrt{a^2-4b}}\Big)}}{24C\sqrt{A^2-4b}}$,
		$h(r)=2\alpha b C^5r^6-3\alpha (10b-a)C^4r^4+$\\
		&$6\alpha(6a+25b-1)C^3r^2+6(\beta+1)C^2r^2$  $\&$ $n=6(4-a-2\beta a+2\beta)C+6\alpha(50b+2a-13a^2-2)C^2$\\
		\hline
		S-T I	\citep{Sunzu18}&$ds^2=-A^2dt^2+\frac{6C}{f(r)}dt^2+r^2d\Omega^2$, where, $f(r)=C^2r^2(6ar^2+3bCr^4+2cC^2r^6+6k)+6C$.\\
		\hline
		S-T II & $ds^2=-A^2(1-aCr^2)^2dt^2-C^2a^3r^2f(r)dr^2+r^2d\Omega^2$\\
		\citep{Sunzu18}&where, $f(r)=\big(2-aCr^2\big)\frac{a^2b}{5}-abf_1(r)+cf_2(r)-a^3C\big((1+f_3(r)\big)$\\
		& $f_1(r)=\frac{1}{4}[\frac{-3}{10}-\frac{3}{5}aCr^2+\frac{1}{2}a^2C^2r^4]$, $f_2(r)=\frac{1}{11}[\frac{1}{5}+\frac{2}{5}aCr^2+a^2C^2r^4-a^3C^3r^6]$ $\&$ $f_3(r)=(1+\frac{k}{(1-3aCr^2)^{2/3}})$.\\
		\hline
		M-K I &$ds^2=-A^2 k^2(aCr^2-1)^{2n_1}e^{\frac{n_2C^4r^8+n_3C^3r^6+n_4C^2r^4+n_5Cr^2+n_6}{6a^5(aCr^2-1)}}dt^2+\frac{dr^2}{(1-aCr^2)^2}+r^2d\Omega^2$\\
		\citep{Malaver20}&with, $n_1=-\frac{(a^2(10a^4-8a^3-2a^2-6ab-6b)-4b^2)\alpha C+(5a^3+a^2+2b)a^2\beta+a^2(a^3+a^2+2b)}{4a^5}$\\
		&$n_2=a^4b^2\alpha C$, $n_3=(15a^3+3a^2+2b)a^3b\alpha C$, \\
		&$n_4=(75a^6+30a^5+3a^4+9a^3b+9a^2b+6b^2)a^2\alpha C-3a^4b(\beta+1)$,\\
		&$n_5=-(75a^6+30 a^5+3 a^4+24a^3b+12a^2b+9b^2)a\alpha C+3a^3b(\beta+1)$ $\&$ \\
		&$n_6=(-3a^6+6a^5-3a^4+6a^3b-6a^2b-3b^2)\alpha C+3a^2(a^2-a^3+b)(\beta+1)$\\
		\hline
		M-K II &$ds^2=-A^2 k^2(2aCr^2-1)^{2n_1}e^{\frac{n_2C^4r^8+n_3C^3r^6+n_4C^2r^4+n_5Cr^2+n_6}{192a^5(aCr^2-1)}}dt^2+\frac{1}{(1-2aCr^2)^2}dr^2+r^2d\Omega^2$\\
		\citep{Malaver20}&with, $n_1=-\frac{(a^2(160a^4-32a^3-2a^2-12ab-3b)-b^2)\alpha C+2(20a^3+a^2+b)a^2\beta+2a^2(4a^3+a^2+b)}{32a^5}$\\
		&$n_2=16a^4b^2\alpha C$, $n_3=16(60a^3+3a^2+b)a^3b\alpha C$, \ $n_4=24(800a^6+80a^5+2a^4+12a^3b+3a^2b+b^2)a^2\alpha C-48a^4b(\beta+1)$,\\
		&$n_5=-6(1600a^6+160 a^5+4 a^4+64a^3b+8a^2b+3b^2)a\alpha C+24a^3b(\beta+1)$ $\&$ \\
		&$n_6=3(-64a^6+32a^5-4a^4+16a^3b-4a^2b-b^2)\alpha C+12a^2(2a^2-8a^3+b)(\beta+1)+\frac{48a^4\gamma}{C}$\\
		\hline
		T-B-G-M	&$ds^2=-f(r)dt^2+\frac{(1-KCr^2)}{(1-Cr^2)}dr^2+r^2d\Omega^2$, \\
		\citep{thiru21}& where, $f(r)=D(1-Cr^2)^{n_1}(1-KCr^2)^{n_2}e^{g(r)}$, 
		$n_1=\left[\frac{\alpha C (3-K)^2}{4(K-1)}-\frac{\beta (3-K)}{4}+\frac{(K-1)(C-\gamma)}{4C}\right]$, \ $n_2=\left[\frac{\alpha C (3-K)^2}{4(K-1)}+\frac{\beta}{2}\right]$\\
		& $\&$, $g(r)=\left[{\frac{\alpha C(K-1)}{2(1-KCr^2)^2}-\frac{\alpha C(2-K)}{1-KCr^2}-\frac{\gamma KCr^2}{4C}}\right]$\\
		\hline\rule[-1ex]{0pt}{3.5ex}
	\end{tabular}
\end{sidewaystable}

\textbf{Working Rule:}\\
\textbf{Thirukkanesh et al.} \citep{thiru21}: In this work, uncharged anisotropic matter with a quadratic equation of state has been considered. Because of the presence of five independent variables $(\rho, p_r, p_t \ \text{or} \ \Delta,\lambda,\nu)$ and only four independent equations (field equations along with $p_r$ as a quadratic equation of $\rho$), they specified one of the gravitational potential $\lambda$ as, $e^{\lambda}=\sqrt{\frac{(1-KCr^2)}{(1-Cr^2)}}$, where $K$ and $C$ are real constants. This form of gravitational potential was originally proposed by Vaidya and Tikekar \citep{Vaidya} to study super-dense stars. Substituting this value for gravitational potential, they have obtained the remaining quantities which are physically reasonable. As the lower central densities and pressures expected for such systems could be less suited to their gravitational model, by avoiding stars of lower mass they have selected  Cen X-3, Vela X-1, and PSR J1614-2230, with evenly spaced masses above $1.4 M_{\odot}$ to study their model. According to the sound speed stability criterion, these models are found stable except possibly near the surface.  However, the authors claimed that this unstable nature of models near the surface may be reduced by increasing the effect of the quadratic term in the EoS used. \\
\textbf{M-K} \citep{Malaver20}: They have generated a physically valid class of charged anisotropic matter with quadratic equation of state in a static spherically symmetric space-time using a gravitational potential $\lambda (r)=-\log(1-\eta aCr^2)$ which depends on an adjustable parameter $\eta$, and the electric field intensity $E=Cr\sqrt{2(a+bCr^2)}$. In this paper only the cases $\eta=1 \ \text{and} \ 2$ has been considered. The MIT bag model can be recovered as a particular case of this work by taking $\beta=0 \ \text{and} \ \gamma \ne 0$.\\
\textbf{Sunzu $\&$ Thomas} \citep{Sunzu18}: Considering the neutral anisotropic fluid matter in static spherically symmetric space-time and using the quadratic equation of state they have generated two classes of exact solutions. In their model, they made choices for $\nu$ and $\Delta$ as, $e^\nu=A\frac{(1-k_1C^mr^{2m})}{(1+k_2C^nr^{2n})}$ and $\Delta=aCr^2+bC^2r^4+cC^3r^6$.  As first case, they have considered $m=n=1 \ \text{and} \ k_1=k_2= 0$. In second one, to obtain different exact solutions they choose $m=n=1, k_1 \ne 0 \ \text{and} \ k_2 = 0$, so that metric function $\nu$ does not remains a constant.\\
\textbf{Malaver} \citep{Malaver17}: In this paper, the author has studied the behavior of relativistic  charged anisotropic 
objects considering Tolman VII form \citep{Tolman39} for the gravitational potential $\lambda$ which is $e^{-\lambda}=\sqrt{1- aCr^2 + bC^2r^4}$ and a quadratic equation of state. Also, he has considered the electrical field as $E=\sqrt{2C(aCr^2-bC^2r^4)}$. This paper is a charged generalization of B-S-P \citep{bhar16}.\\
\textbf{Bhar et al.} \citep{bhar16}: In this article, the authors have presented a well-behaved solution by considering the quadratic equation of state for the neutral anisotropic matter distribution for the first time in Tolman VII background. Using this obtained solution, they have optimized the masses and radii of compact stars  RXJ1856.5-3754, Her X-1, Cyg X-2, PSR B0943+10, and PSR B1913+16, and compared it with their experimentally observed values.\\
\textbf{Ngubelanga et al.} \citep{ngu15b}: In this paper, the authors have found new exact solutions to the Einstein-Maxwell field equations for matter configurations with anisotropy and charge in isotropic coordinates. They have selected the barotropic equation of state to be quadratic and chosen electric field $E=r\sqrt{(C + Dr^2)}$. They have regained the masses of compact stars  4U 1608-52, PSR J1614-2230, PSR J1903+0327, EXO 1745-248, and SAX J1808.4-3658, using particular choices of parameters. To show the model realistic, they have done a comprehensive physical analysis for the star PSR J1903+0327.
\clearpage
\subsection{Colour-flavour-locked equation of states} 
Colour–flavor locking is a phenomenon that is normally expected to occur in ultra-high-density strange matter, a form of quark matter. The quarks form Cooper pairs, whose colour properties are correlated with their flavor properties in a one-to-one correspondence between three colour pairs and three flavor pairs. At sufficiently large baryon chemical potential, the more stable configuration of strange quark matter (SQM) is the colour flavor locked (CFL) \citep{alford} superconducting phase. In this phase, Cooper pairs of quarks of different colours and flavors are coupled with total zero momentum. For a very high density, the mass of the strange quark is negligible in comparison to the baryonic chemical potential, which leads to the same density of the three flavors of up $(u)$, down $(d)$ and strange $(s)$ quarks \citep{Orsaria}. As a result, the CFL phase is naturally electrically neutral.

By appealing to the microphysics associated with strange matter, Rocha et al. \citep{Rocha19} and many others adopted a CFL EoS of the form 
\begin{equation}
	\nonumber	p_r=\alpha \rho+\frac{\beta}{\sqrt{\rho}}-\gamma
\end{equation}
which generalises the MIT Bag model and describes the internal structure of a compact star made of strange matter in the colour-flavor-locked phase. When $\beta= 0$ the CFL EoS becomes a linear EoS.

Edward Witten was one of the first to propose, in 1984, that the strange matter should be the true ground state of hadrons, instead of ${}^{56}$Fe, having a lower energy per baryon than ordinary nuclei. This matter is assumed to be composed of roughly equal numbers of u, d, and s quarks, and a small number of electrons to attain charge neutrality. If this hypothesis is correct, neutron stars would be strange stars, or atleast hybrid stars with a thin crust of nuclei, where the temperature and pressure conditions are extreme enough to convert hadronic matter into this new stable phase of quarks. Great effort has been made to describe correctly and accurately the physical features of these compact objects since Witten’s idea. Alcock et al. \citep{Aclock} first described a strange star using the MIT bag model with a linear EoS that did not include the strange quark mass and assumed quarks as asymptotically free particles. In the simplest approach, the CFL EoS is obtained by starting with the thermodynamic potential, constructed in detail in references \citep{Raja01,alf01}. Lugones and Horvath \citep{Lugones} proposed a model, where they assume that the CFL phase is the real ground state of matter since the Fermi energy of the system is lowered by the formation of Cooper pairs. Later work showed that a paired, symmetric in flavor, the colour-flavor locked state would be preferred to the one without any pairing for a wide range of the parameters. Recently, Singh et al. \citep{singh21} presented an exact solution that could describe compact stars composed of colour-flavor locked (CFL) phase. This work shows that the anisotropy of the pressure at the interior increases with the colour superconducting gap and it leads to a decrease in the adiabatic index closer to the critical limit. Also, the fluctuating range of mass due to the density perturbation in their model is larger for the lower colour superconducting gap, which results in a more stable configuration.

In Table \ref{tc} we have given some metrics for anisotropic stellar models generated by a CFL EoS. We have also briefly discussed these solutions.

\begin{sidewaystable}[h]
	\centering
	\caption{\label{tc} Solutions generated by CFL EoS.}
	\begin{tabular}{|l|l|}
		\hline
		name [ref.] & metric  \\
		\hline
		R-B-A-H I	&$ds^2=-c^2A^2e^{f}(aCr^2-1)^{-(1+5\alpha)/2}dt^2+(1-aCr^2)^{-n}dr^2+r^2d\Omega^2$\\
		\citep{Rocha20}	&$f=\frac{-a C(1+\alpha)+\gamma-\sqrt{aC(6-5aCr^2)}}{2a C(a Cr^2-1)}+\frac{5\beta \sqrt{a C}\arctan(\sqrt{6-5aCr^2})}{2a C}$, $\alpha=\frac{1}{3}$, $\beta=\frac{2\eta}{\pi}$ $\&$ $\gamma=\frac{3\eta^2}{\pi^2}+\frac{4}{3}B$ \\
		&with $\eta=-\frac{m_s^2}{6}+\frac{2\Delta^2}{3}$, where $m_s$  is the mass of quark s.\\
		\hline
		R-B-A-H II	&$ds^2=-c^2Ke^{2W^2}(1+ar^2)^{1/3}(1+(a-b)r^2)^{Y}dt^2+\frac{1+ar^2}{1+(a-b)r^2}dr^2+r^2d\Omega^2$\\
		\citep{Rocha20}	&where, $K$, $W$ and $Y$ are functions of $\alpha$, $\beta$ and $\gamma$,  $\alpha=\frac{1}{3}$, $\beta=\frac{2\eta}{\pi}$ $\&$ $\gamma=\frac{3\eta^2}{\pi^2}+\frac{4}{3}B$ \\
		&with $\eta=-\frac{m_s^2}{6}+\frac{2\Delta^2}{3}$, $m_s$  being the mass of quark s.\\
		&\\
		\hline
		T-K-G I \citep{thiru20}&$ds^2=-d^2\left(\frac{1+c^2r^2}{1-c^2r^2}\right)^{\frac{\mu(1+3\alpha)-3(\gamma+\beta \sqrt{\mu})}{\mu}}dt^2+\frac{12c^2}{\mu(1+c^2r^2)^2}(dr^2+r^2d\Omega^2)$\\
		\hline
		T-K-G II\citep{thiru20}&$ds^2=-d^2\left(1+br^2\right)^{\frac{1+5\alpha}{2}}\left(\sqrt{1+br^2}+\sqrt{6+br^2}\right)^{\frac{-5\alpha \beta}{2\sqrt{b}}}e^fdt^2+\frac{a^2}{1+c^2r^2}(dr^2+r^2d\Omega^2)$\\
		&where, $f=2\left(b(1+\alpha)-a^2\gamma\right)(1+br^2)-\frac{\alpha \beta}{2\sqrt{b}}\sqrt{1+br^2}\sqrt{6+br^2}$\\
		\hline
	\end{tabular}
\end{sidewaystable} 

\textbf{Working Rule :}\\
\textbf{Rocha et al.} \citep{Rocha19,Rocha20}: The authors have provided two anisotropic models made of CFL strange matter. The strange matter is assumed to be composed of strange quarks in addition to the usual ups and downs, having an energy per baryon lower than the strangeness counterpart, and even lower than that of nuclear matter. With CFL EoS ($\alpha=\frac{1}{3}$, $\beta=\frac{2\eta}{\pi}$ $\&$ $\gamma=\frac{3\eta^2}{\pi^2}+\frac{4}{3}B$ with $\eta=-\frac{m_s^2}{6}+\frac{2\Delta^2}{3}$, where $m_s$  is the mass of quark s.),  they have used Thirukkanesh and Ragel ansatz and Sharma-Maharaj ansatz for their first and second model respectively. All present observational bounds on achieved masses and radii are satisfied by their models. According to them, some additional requirements being evaluated (coming from X-ray burst light curves, Quasi-periodic oscillations, and other phenomena observed) can be addressed using the obtained exact expressions.\\
\textbf{Thirukkanesh et al.} \citep{thiru20}: In this study, authors have modeled spherically symmetric compact objects by adopting an isotropic line element which is simultaneously comoving. The interior matter configuration is described by an anisotropic fluid with unequal stresses in the radial and tangential directions.  They have invoked the Durgapal and Banerjee transformation and a simple ansatz for one of the gravitational potentials to obtain an exact solution describing a compact object obeying a CFL EoS. The CFL EoS allows integrating the field equations thus reducing the problem to a single generating function of one of the gravitational potentials. By appealing to the regularity, stability, and physical viability of the gravitational and thermodynamical variables required for a realistic description of compact objects they have presented a tractable stellar model. They have examined the behavior of the CFL model in the linear approximation and show that the more robust linear EoS mimics the CFL EoS to a very good approximation. Their study focused on the comparison between the CFL and linear EoS exact solutions for compact stars consisting of the strange matter phase. By plotting the various thermodynamical quantities, stability criteria, and energy conditions they have shown that the linear EoS gives a reasonable approximation to the CFL EoS.
\clearpage
\subsection{Polytropic equation of state}
Polytropes are very useful to study the internal structure of stars. They can be used to discuss various astronomical situations and descriptions of compact objects. The
polytropic models have gained much attention in recent times because of their elementary equation of state (EoS). Chandrasekhar \citep{Chandra1939} developed the basic theory of Newtonian polytropes by using laws of thermodynamics for polytropic spheres. Irregularities arise in Chandrasekhar’s theory of slowly rotating polytropes were removed by Kovetz \citep{Kovetz}. Discussion of white dwarfs in the perspective of polytropes is made in \citep{Shapiro}, while Abramowicz \citep{Abramowicz} introduced the concept of higher-dimensional polytropes. Komatsu et al. \citep{Komatsu}] explored the applications of rapidly rotating gravitating sources to uniformly rotating polytropes by using numerical techniques. Cook et al. \citep{Cook} determined the maximum mass and spin rate to diagnose stability for rotating polytropes. Azam et al. \citep{Azam1,Azam2,Azam3} constructed a general framework to study the physical attributes of charged cylindrical and spherical sources via generalized polytropic EoS. Isayev \citep{Isayev} discussed general relativistic polytropes in anisotropic stars.

The generalized polytropic gas model for stellar like structure assumes an equation of state of the form 
\begin{equation}
	\nonumber	p_r = \beta \rho +\alpha\rho^{1+\frac{1}{\eta}}
\end{equation}
which is combination of linear EoS ($p_r=\beta\rho$) and polytropic EoS ($p_r= \alpha \rho^{1+\frac{1}{\eta}}$).
Here, $\eta$ is the polytropic index, which defines intermediate cases between isothermic and adiabatic thermodynamical processes, and $\beta$ is the polytropic constant. 

The physical radius and mass of the spherical body appear only in the polytropic constant, and hence, the stability of a star is independent of its size or mass, and different types of stars correspond to different polytropic indices $\eta$. The first solar model ever considered, developed by Eddington in 1926, was that of an $\eta$ = 3 polytropic star. Although somewhat incomplete, this simplified model gives rise to relevant constraints on the physical quantities. Various developments have been made in the recent past to explain stellar structure either by using Schwarzschild or isotropic coordinates for polytropic index ranging from $1 < \eta < 5$. Pandey et al. \citep{Pandey} explained properties of polytropic models with index $\frac{1}{2} < \eta < 3$, different values of polytropic indices have been chosen in \citep{Thirukkanesh12} to study the effect of an anisotropic factor on spherically symmetric exact solutions. Ngubelanga and Maharaj \citep{ngu15a,Ngu17} found new classes of polytropic models and generate physically acceptable models for different values of polytropic indices. 

Polytropes and their stability have been studied by Chandrasekhar who worked out the requirement for the stability of a self-gravitating sphere. Solutions to the Einstein field equations admitting a polytropic equation of state are not common with physical viable solutions being a rarity. For recent works on anisotropic stellar modeling by employing polytropic EoS one can go through the papers \citep{Isayev,nasim,Azam3,Noureen,Mardan}. We have enlisted some metrics in Table \ref{tp} which have been used to study the anisotropic stellar modeling with polytropic EoS and discussed the solutions briefly.

\begin{sidewaystable}[h]
	\centering
	\caption{\label{tp} Solutions generated with polytropic EoS.} 
	\tiny	\begin{tabular}{|l|l|}
		\hline
		name [ref.] & metric  \\
		\hline
		N-M-A-S-K I&$ds^2=-\left[\frac{K(a-br^2)^{n_1}(a+br^2)^{n_2}\exp{\frac{-d^2 \alpha r^2}{128 b^2 \pi}}}{br^2+a}\right]^2dt^2+\frac{1}{(br^2+a)^2}[dr^2+r^2d\Omega^2]$\\
		\citep{Noureen}&where, $n_1=\frac{1}{256 \pi ab^3}[\alpha m^2+32 \pi b \beta m+16 \pi b]$, $m=(24ab^2-bc+ad)$ and\\
		& $n_2=\frac{1}{256 \pi ab^3}[16 \pi b(ad+bc-8ab^2)-32 \pi b \beta m-\alpha m^2]$\\
		\hline
		N-M-A-S-K II&$ds^2=-\left[\frac{K(a-br^2)^{n_1}(a+br^2)^{n_2}[g(r)]^{n_3}[g^{-1}(r)]^{n_4}\exp{\frac{d\alpha }{16b^2}\sqrt{\frac{f(r)}{\pi}}}}{br^2+a}\right]^2dt^2+\frac{1}{(br^2+a)^2}[dr^2+r^2d\Omega^2]$;\\
		\citep{Noureen}	& $n_1=\frac{1}{16 ab^2}[2\beta m-(ad+bc-8ab^2)]$, $m=(24ab^2-bc+ad)$, $f(r)=24 ab-c-dr^2$, \\
		& $n_2=\frac{m}{16 ab^2}(2\beta +1)$, $n_3=\frac{\alpha}{64a}\sqrt{\frac{(m-2ad)^3}{\pi b^5}}$, $n_4=\frac{\alpha}{64a}\sqrt{\frac{m^3}{\pi b^5}}$ and $g(r)=\frac{\sqrt{m-2ad}+\sqrt{bf(r)}}{\sqrt{m-2ad}-\sqrt{bf(r)}}$.\\
		\hline
		N-M-A-S-K III&$ds^2=-\left[\frac{K(a-br^2)^{n_1}(a+br^2)^{n_2}g(r)g^{-1}(r)\exp{\frac{d\alpha}{768
					b^2}\sqrt{\frac{f(r)}{\pi^3}}(168ab-7c-dr^2)}}{br^2+a}\right]^2dt^2+\frac{1}{(br^2+a)^2}[dr^2+r^2d\Omega^2]$\\
		\citep{Noureen}	&where, $n_1=\frac{1}{16ab^2}[2\beta(m-2ad)-(ad+bc-8ab^2)]$, $m=(24ab^2-bc+ad)$, $n_2=\frac{m}{16ab^2}(2\beta+1)$,\\
		&  $n_3=\frac{\alpha}{1024a}\sqrt{\frac{(m-2ad)^5}{\pi^3 b^7}}$, $n_4=\frac{\alpha}{1024a}\sqrt{\frac{m^5}{\pi^3 b^7}}$, $g(r)=\frac{\sqrt{m-2ad}+\sqrt{bf(r)}}{\sqrt{m-2ad}-\sqrt{bf(r)}}$ and  $f(r)=24ab-c-dr^2$.\\
		\hline
		N-M-A-S-K IV&$ds^2=-\left[\frac{K(a-br^2)^{n_1}(a+br^2)^{n_2}\exp{\frac{d^2 \alpha [6c-144ab+dr^2]r^2}{4096 b^2 \pi^2}}}{br^2+a}\right]^2dt^2+\frac{1}{(br^2+a)^2}[dr^2+r^2d\Omega^2]$\\
		\citep{Noureen}	&where, $n_1=\frac{1}{4096 \pi^2 ab^4}[\alpha m_1+256 \pi^2 b^2 m_2+512 \pi^2 b^2 \beta m_2]$, $m_1=(24ab^2-bc+ad)$, $m_2=(-8ab^2+bc+ad)$\\
		&and $n_2=\frac{1}{4096 \pi^2 ab^4}[\alpha {m_1}^3+256\pi^2 b^2 m_1-512\pi^2 b^2 \beta\alpha m_1]$\\
		\hline
		M-R-N-J I&$ds^2=-K(a-br^2)^{2n_1}(a+br^2)^{2(n_2-1)}\exp{\frac{-16 d^2 \alpha \pi^3 r^2}{b^2}}dt^2+(br^2+a)^{-2}[dr^2+r^2d\Omega^2]$\\
		\citep{Mardan}	&where, $n_1=\frac{1}{8 \pi ab^3}[2\alpha m_1+ \pi b m_2+b \pi \beta m_3]$, $m_1=(-3ab^2+4ad\pi^2+4bc\pi^2+2b\pi \epsilon)^2$,\\
		&$m_2=(ab^2-16bc\pi^2-16ad\pi^2+8b\pi \epsilon)$, $m_3=12ab^2-16ad\pi^2-16bc\pi^2-8b\epsilon \pi$ \\
		&$n_2=\frac{1}{8 \pi ab^3}[2\alpha m_4+ \pi b m_5+b \pi \beta m_6]$, $m_4=(3ab^2+4ad\pi^2-4bc\pi^2-2b\pi \epsilon)^2$, $m_5=(3ab^2-16bc\pi^2+16ad\pi^2+$ \\
		&$8b\pi \epsilon)$ and $m_6=12ab^2+16ad\pi^2-16bc\pi^2-8b\epsilon \pi$. \\
		\hline
		M-R-N-J II&$ds^2=-K(a-br^2)^{2n_1}(a+br^2)^{2(n_2-1)}\exp{\frac{8d^2 \alpha \pi^2 r^2C}{b^2}}dt^2+(br^2+a)^{-2}[dr^2+r^2d\Omega^2]$\\
		\citep{Mardan}	&where, $C=(-9ab+12c\pi^2+2dr^2\pi^2+6\pi \epsilon)$, $n_1=\frac{1}{8 \pi^2 ab^4}[\alpha m_1^3-\pi b m_2-b \pi \beta m_3]$, $m_1=-3ab^2+4ad\pi^2+$\\
		&$4bc\pi^2+2b\pi \epsilon$, $m_2=(ab^2-16bc\pi^2-16ad\pi^2+8b\pi \epsilon)$, $m_3=12ab^2-16ad\pi^2-16bc\pi^2-8b\epsilon \pi$,  \\
		& $n_2=\frac{1}{8 \pi^2 ab^4}[\alpha m_4^3+ \pi^2 b^2 (m_5+\beta m_6)]$, $m_5=(3ab^2-16bc\pi^2+16ad\pi^2+8b\pi \epsilon)$ and\\
		& $m_6=12ab^2+16ad\pi^2-16bc\pi^2-8b\epsilon \pi$. \\
		\hline
		M-R-N-J III&$ds^2=-K(a-br^2)^{2n_1}(a+br^2)^{2(n_2-1)}[g_1(r)]^{n_3}[g_2(r)]^{2n_4}\exp{\frac{8d\alpha \pi^{3/2} }{b^2}\sqrt{2f(r)}}dt^2+(br^2+a)^{-2}[dr^2+r^2d\Omega^2]$;\\
		\citep{Mardan}	& $n_1=\frac{1}{8 \pi^2 ab^4}[a(b^2-16d\pi^2)+8\pi b(-2c\pi+\epsilon)+4\beta(s_1-s_2)]$, $f(r)=3ab-4\pi^2c+4\pi^2dr^2+\epsilon$, \\
		& $n_2=\frac{1}{8 \pi^2 ab^4}[a(3b^2+16d\pi^2)+8\pi b(-2c\pi+\epsilon)+4\beta(s_1-s_2)]$, $n_3=\frac{\alpha (s_1-s_2^{3/2})}{2ab^5}$, $n_4=\frac{-\alpha (s_3-s_2^{3/2})}{2ab^{5}}$, \\ 
		&$g_1(r)=\frac{\sqrt{s1+s_2}+\sqrt{bf(r)}}{\sqrt{s1-s_2}-\sqrt{bf(r)}}$,
		$g_2(r)=\frac{\sqrt{s_3+s_2}+\sqrt{bf(r)}}{\sqrt{s3-s_2}-\sqrt{bf(r)}}$, $s_1=a(3b^2-4d\pi^2)$, $s_2=2b\pi(2\pi c+\epsilon)$ and $s_3=a(3b^2+4d\pi^2)$.\\
		\hline
	\end{tabular}
\end{sidewaystable}

\textbf{Working rule:}\\
\textbf{Noureen et al.} \citep{Noureen}: In this work, authors have studied the combined effect of charge and anisotropy on gravitational interaction of compact sources by making use of the generalized polytropic equation of state. They have discussed four different values of the polytropic index ($\eta$=1,2,$\frac{2}{3},\frac{1}{2}$) to ascertain the solution of Einstein-Maxwell field equations and develop a new class of spherically symmetric charged polytropic models. They made a physically acceptable choice for the electric field as $E=\sqrt{c+dr^2}$ and one metric potential as $\frac{1}{(br^2+a)^2}$, with constants a, b, c and d. They have regained the masses of realistic strange stars 4U 1820-30, PSR J1614-2230, PSR J1903+327, Vela 4U, and Vela X-1 that show the viability of the model. \\
\textbf{Mardan et al.} \citep{Mardan}: In this paper, generalized polytropic equation of state is used to generate well behaved and physically viable new classes of polytropic models from the solution of Einstein-Maxwell field equations for charged anisotropic fluid configuration. They have discussed four different values of the polytropic index ($\eta$=1, $\frac{1}{2}$, 2) to ascertain the solution. They choose an electric field as $E=\sqrt{c+dr^2}$ and one metric potential as $\frac{1}{(br^2+a)^2}$, with constants a, b, c, and d. more general and singularity free. (i) if ``$E = 0, \epsilon= 0$” is substituted, the models reduces to the case presented by Mardan et al. [11], (ii) if ``$\epsilon = 0$” is placed, the models developed by Noureen et al. \citep{Noureen} is obtained, (iii) by ``$E = 0$” the models given by Mardan et al. [25] can be generated.\\

\clearpage
\subsection{Chaplygin gas model EoS}
The proposal regarding treating dark energy and dark matter as different manifestations of a single entity leads to the Chaplygin gas model. In the Chaplygin gas model, the equation of state derives from string theory. The Chaplygin gas model is one of the predicting candidates used to explain the present accelerated expansion of the universe. Apart from its connection to the string theory, an additional attractive feature of this model is that such an approach naturally solves, at least phenomenologically, the coincidence problem. The Chaplygin gas model first 
represents the dust behavior and in late times as cosmological constant. Various applications of the Chaplygin gas model have been pursued to account for cosmological observations such as acceleration of the cosmic fluid and structure formation. The Chaplygin equation of state has been subsequently modified to a more generalized Chaplygin gas EoS,  which depicts a mixture of the cosmological constant and a perfect fluid with an equation of state, $p=-\frac{K}{\rho^{\alpha}}$ in intermediate stages. The parameter involves in this equation, makes the Chaplygin gas model more flexible. 

The equation of state which combines many models of Chaplygin gas model is given by
\begin{equation}
	p=H \rho-\frac{K}{\rho^{\alpha}}
\end{equation}
where $\alpha$, H and K are constants with $K>0$ and $\alpha \ge 0$.\\

Recently, scientists have shown great interest in Chaplygin gas EoS to explain the accelerating phase of the present Universe as well as to unify the dark energy and dark matter. As Chaplygin gas EOS is a specific form of polytropic EOS of negative index $n =-1/(1+\alpha)$, so it looks promising to describe dark energy spherically symmetric charged objects, generally termed as dark stars in the literature. The generalized Chaplygin equation of state has been employed by many researchers to model dark stars which are remnants of continued gravitational collapse. 

The following Table \ref{tch} gives the name and reference of the studied exact solutions of the Einstein field equations using the Chaplygin Gas Model EoS, along with the corresponding metric.

\begin{sidewaystable}[h]
	\centering
	\caption{\label{tch} Solutions obtained through Chaplygin gas model EoS.}
	\tiny	\begin{tabular}{|l|l|}
		\hline
		$name [ref.]$ & $metric$  \\
		\hline
		
		R-R-J-C \citep{rah10} &  $ds^2=-e^{br^2 +c}dt^2+e^{ar^2}dr^2+r^2 d\Omega^2$\\
		\hline
		Bhar \citep{bhar15}&$ds^2=-e^{2f(r)}dt^2+(1+\frac{r^2}{R^2})dr^2+r^2 d\Omega^2$\\
		&$f(r)=\frac{\alpha}{4}[\frac{r^2}{R^2}+2\log(1+\frac{r^2}{R^2})]+\frac{(8\pi rR)^2\beta}{4}[\frac{1}{3}(3+\frac{r^2}{R^2})^3-3(3+\frac{r^2}{R^2})^2]+\frac{(8\pi R^2)^2\beta}{4}[\frac{12r^2}{R^2}-8\log(3+\frac{r^2}{R^2})]+C$\\
		\hline
		B-G-S \citep{bhar18}& $ds^2=-f^2(r) dt^2+\frac{a^2}{1+br^2}[dr^2+r^2 d\Omega^2]$, with \\
		&$f(r)=c\exp{\big[(1+br^2)\big(\frac{1+H}{4}-\frac{K\kappa^2a^4}{4b^2}\big)\big]}(1+br^2)^{(1+5H)/4}(6+br^2)^{5K\kappa^2a^4/{4b^2}}$\\
		\hline
		O-M-R-L \citep{Ortiz0}&$ds^2=-[F^2(Cr^2+3)^\frac{K}{2C^2}(Cr^2+1)^H(Cr^2-1)^{-2H-1\frac{K}{2C^2}}e^{\frac{Kr^2(Cr^2+2)}{8C}}]dt^2+\frac{1+Cr^2}{1-Cr^2}dr^2+r^2 d\Omega^2$\\
		\hline
		P-K-S& $ds^2=e^{2f(r)}dt^2-\frac{K_1(1+Cr^2)}{K_1+Cr^2}dr^2-r^2 d\Omega^2$\\
		\citep{amit}	&$f(r)=\frac{-H}{4}\left[(3-K_1)\log{K_1+Cr^2}\right]-\frac{\kappa^2 KK_1}{4C(K_1-1)}\left[\frac{Cr^4}{2}-Kr^2-\frac{8\log{C+3r^2}}{C(K_1-3)}-\frac{(K_1-1)^3\log{(K_1+Cr^2)}}{C(K_1-3)}\right]+$\\
		&$\frac{(K_1-1)\log{K_1+Cr^2}}{4}+D$\\
		\hline
		M-K \citep{Mala}& $ds^2=-[c_1^2H^2(bC^2r^4+aCr^2-3a)^{2n_1}(aCr^2-1)^{2n_2}e^{2(n_3C^2r^4+n_4Cr^2+2Ka^2(a^2+2b)\tan^{-1}\frac{2bCr^2+a}{\sqrt{a(a+12b)}})}]dt^2+$\\
		&$(1-aCr^2)dr^2+r^2 d\Omega^2$; with $n_1=-\frac{aK}{8C^2(3a^3-a^2-b)}$, $n_2=\frac{3(3H+1)a^6-2(3H+2)a^5+(H-K+1)a^4-2(3H+2)ba^3+(H+1)(2ba^3+b^2)}{4a^3(3a^3-a^2-b)}$,\\
		&$n_3=\frac{(H+1)(3a^2b-ab-\frac{b^2}{a})}{8(3a^3-a^2-b)}$ and $n_4=\frac{2(H+1)(3a^3-a^2+3ab-2b-\frac{b^2}{a})}{8(3a^3-a^2-b)}$.\\
		\hline
	\end{tabular}
\end{sidewaystable}

\textbf{Working Rule:}\\
\textbf{Rahman et al.} \citep{rah10}: To obtain a non singular, well behaved and stable model, for charged fluid sources with anisotropic stresses, the authors have considered Krori-Barua \citep{kb} ansatz: $\lambda(r) = ar^2$ and $\nu(r) = br^2 + c$, where, a, b and c are some arbitrary constants. These constants are determined in terms of the physical quantities mass, charge, and radius of the source. Along with this they have considered the non-linear Chaplygin gas EOS and assumed $\alpha=1$ and $H\ne-1$. They fixed the coefficients of Chaplygin gas EOS by the matching conditions at the boundary of the source. \\
\textbf{Bhar} \citep{bhar15} : In this model the author has developed a model of anisotropic strange star by assuming Finch-skea ansatz $e^\lambda=\sqrt{1+\frac{r^2}{R^2}}$, which admits Chaplygin equation of state. He has shown the consistency of the model with the observational data of strange star SAX J 1808.4-3658.\\
\textbf{Bhar et al.} \citep{bhar18}: The authors have presented a model describing a static spherically symmetric anisotropic fluid configuration, in which the matter content obeys a generalized Chaplygin equation of state. They have assumed $\alpha=1$ and taken one of the gravitational potentials as $\frac{a^2}{1+br^2}$, where a and b are constants. To investigate the relevance of the model in the studies of compact stars, they have considered compact stars, namely 4U 1538-52, PSR J1614-2230, Vela X-1, and Cen X-3, and showed that the estimated radii and masses determined from our model are very close the observed values of these stars and demonstrated that their solution can be used as a realistic model for describing ultracompact stars within the framework of general relativity.\\
\textbf{Ortiz et al.} \citep{Ortiz0}: In this paper, an uncharged anisotropic fluid model has been obtained utilizing the generalized Chaplygin equation of state with $\alpha=1$ and imposing a particular form of metric potential as $e^\lambda=\sqrt{\frac{1+Cr^2}{1-Cr^2}}$. To confirm the feasibility of the model, the authors have performed a complete physical analysis. They have shown the stability of the system through the pressure wave velocities and the relativistic adiabatic index. To mimic a realistic compact object, they have imposed the radius to be 9.5 (Km). The resulting numerical values for the fluid characteristics of the model indicate that the structure could represent a quark star mixed with dark energy.\\
\textbf{Prasad et al.} \citep{amit}: In this paper, the authors have derived a new model for anisotropic compact stars with the modified Chaplygin equation of state, coupled with a suitable form of gravitational potential namely Buchdahl ansatz. They have discussed the stability and feasibility of the solution. Their solution is continuous and maintains hydrostatic equilibrium in the interior of the star. They have verified that the stars like PSR B0943+10 for $K_1 = -1.52$, Her X-1 for $K_1 =-11.52$ and SAX J1808.4-3658 for $K_1 = 14.50$ to be close candidates for this model. \\
\textbf{Malaver $\&$ Kasmaei} \citep{Mala}: Considering the Chaplygin equation of state with $\alpha=1$, in this work, the authors have found a completely new model for a compact star with charged anisotropic matter distribution. To construct the model they considered the metric potential and electric field intensity in the form of  $e^\lambda=\sqrt{\frac{1}{1-aCr^2}}$ and $E=\sqrt{2C^2r^4(a+bCr^2)}$. The obtained model satisfies all the expected physical properties for a realistic compact star. The authors have claimed that the solution obtained in this work can have multiple applications in astrophysics and cosmology.
\clearpage
\subsection{Van der Waals EoS}
The Van der Waals (VDW) equation of state has fascinated researchers for a long time, largely because of its flexibility. 
The Van der Waals  EoS was presented by Van der Waals in his 1873 Ph.D. thesis and for this, he was awarded the Nobel Prize in Physics 1910.  The VDW model is useful to explain the universe with a minimal number of ingredients, and the VDW gas treats dark matter and dark energy as a single fluid.

It was observed that assuming dark energy is governed by a perfect fluid EoS may systematically induce wrong results and be misleading in inferring the nature of dark energy. In this spirit, an alternative model was proposed without the presence of exotic fluids, using a more complicated equation of state, namely, the Van der Waals equation of state \citep{Capoz02,Capoz03},  given by
\begin{equation}
	p_r=\frac{\gamma \rho }{1-\beta \rho}-\alpha \rho^2
\end{equation}
where, $\alpha$, $\beta$, and $\gamma$ are the parameters of the equation of state. The success of this model has stimulated several different approaches.

The relativistic stellar models with a Van der Waals equation of state are studied in the treatment of Lobo \citep{Lobo07}. Lobo \citep{Lobo07} initiated the study of Van der Waals quintessence stars. Apart from variations of $\rho$ and $p_r$ in the interior of the VDW fluid, which may be caused by gravitational instabilities, this author imposes a cut-off of the energy-momentum tensor at r = R, where the internal metric matches the external Schwarzschild solution. The Van Der Waals quintessence EoS is an interesting scenario for describing the late universe and seems to provide a solution to the puzzle of dark energy, without the presence of exotic fluids.

The VDW EoS is further modified by Malaver \citep{Mala13a}. The modified/generalized Van der Waals EoS is given by
\begin{equation}
	p_r=\alpha \rho^{\Gamma+1}+\frac{\gamma \rho^\Gamma}{1+\beta \rho^\Gamma}
\end{equation}
where $\alpha$, $\beta$ and $\gamma$ are arbitrary constants and $\Gamma = 1 +\frac{1}{\eta}$, with $\eta$ is the polytropic index.

Malaver \citep{Mala13a,Mala13b} found charged stellar models with Van der Waals as well as generalized Van der Waals equation of state. 
Thirukkanesh and Ragel \citep{Thiru14} formulate a system of field equations with a modified Van der Waals type EoS to the Einstein field equations in spherically symmetric static spacetime to describe anisotropic compact matter. 
Anisotropic models with Van der Waals modified equation of state also include recent findings of Sunzu and Mahali \citep{sunzu18}, Malaver and Kasmaei \citep{Malaver19}.
In Table \ref{tvdw} we have given some metrics which have been used to study exact solutions of the Einstein field equations using the Van der waals EoS. 

\begin{sidewaystable}[h]
	\centering
	\caption{\label{tvdw} Solutions obtained through Van der Waals EoS.}
	\tiny	\begin{tabular}{|l|l|}
		\hline
		$name [ref.]$ & $metric$  \\
		\hline
		Malaver \citep{Mala13a}&$-A^2y^2(x)dt^2+\frac{(1+\beta x)}{(1+\alpha x)4cx}dx^2+\frac{x}{c}d\Omega^2$ with $x=cr^2$ and $y$ is a solution of $\frac{\dot{y}}{y}+\frac{\alpha-\beta}{4(1+\alpha x)}=\frac{\alpha c [2(\beta-\alpha)(3+\beta x)-k(3+\alpha x)]^2}{4(1+\alpha x)(1+\beta x)^3}+$\\
		&$\frac{\gamma (1+\beta x) [2(\beta-\alpha)(3+\beta x)-k(3+\alpha x)]}{4(1+\alpha x)\{2(1+\beta x)^2+\beta c [2(\beta-\alpha)(3+\beta x)-k(3+\alpha x)]\}}-\frac{k(3+\alpha x)}{8(1+\alpha x)(1=\beta x)}$.\\
		\hline
		Malaver \citep{Mala13b}&$-A^2y^2(x)dt^2+\frac{(1+\alpha x)}{(1+\beta x)4cx}dx^2+\frac{x}{c}d\Omega^2$ with $x=cr^2$ and $y$ is a solution of $\frac{4(1+\beta x)}{1+\alpha x}\frac{\dot{y}}{y}-\frac{\alpha-\beta}{1+\alpha x}=\frac{\alpha [(\alpha-\beta)(3+\alpha x)-\beta x]^{\Gamma+1}}{C(1+\alpha x)^{2(\Gamma+1)}}+$\\
		&$\frac{\beta [(\alpha-\beta)(3+\alpha x)-\beta x]^\Gamma}{C\{(1+\alpha x)^{2\Gamma}+\Gamma [(\alpha-\beta)(3+\alpha x)-\beta x]^\Gamma\}}-\frac{\beta x}{(1+\alpha x)^2}$.\\
		\hline	
		M-K I \citep{Malaver19} & $ds^2=-{A^2{c_1}^2(5a^2\beta c^2 r^2 -12 a \beta c-4)^{2n_1}(acr^2-2)^{2n_2}e^{\frac{n_3c^2r^4+n_4cr^2+n_5}{16(a\beta c+2)(2-acr^2)}}}dt^2+{(1-\frac{1}{2} acr^2)^{-2}}dr^2+r^2 d\Omega^2$\\
		&with $n_1=\frac{5\gamma}{(a\beta c+2)^2}$, $n_2= \frac{-5a^3\beta^2c^3+2a^2\beta^2c^2-20a^2\beta c^2+8a\beta c-20 a c-40\gamma+8}{8(a\beta c+2)^2}$, $n_3=-25a^3c(a\beta c+2)$, $n_4=50a^2c(a\beta c+2)$\\
		&$\&$ $n_5=4a\beta c(ac-4)+8(ac+4\gamma-4)$.\\
		\hline
		M-K II \citep{Malaver19} & $ds^2=-{A^2{c_2}^2(20a^2\beta c^2 r^2 -12 a \beta c-1)^{2n_1}(2acr^2-1)^{2n_2}e^{\frac{n_3c^2r^4+n_4cr^2+n_5}{4(2a\beta c+1)(1-2acr^2)}}}dt^2+{(1-2 acr^2)^{-2}}dr^2+r^2 d\Omega^2$\\
		&with $n_1=\frac{5\gamma}{4(2a\beta c+1)^2}$, $n_2= \frac{-160a^3\beta^2c^3-4a^2\beta^2c^2-160a^2\beta c^2-4a\beta c-40 a c-5\gamma-1}{4(2a\beta c+1)^2}$, $n_3=-400a^3c(2a\beta c+1)$, \\
		&$n_4=200a^2c(2a\beta c+1)$ $\&$ $n_5=2a\beta c(4ac+1)+(4ac+\gamma+1)$.\\
		\hline
	\end{tabular}
\end{sidewaystable}
\textbf{Working Rule:}\\
\textbf{Malaver} \citep{Mala13a}: In this paper Malaver has extended the work of Feroze and Siddiqui \citep{Feroze} by considering Van der Waals modified equation
of state for charged anisotropic matter distributions. He assumed a different gravitational potential and the electrical field which depend on adjustable parameters, given as  $\lambda=\frac{1}{2}(\log(1+\beta cr^2)-\log(1+\alpha cr^2))$ and $E=\sqrt{\frac{ck(3+\alpha cr^2)}{(1+\beta cr^2)^2}}$. The obtained new solutions of the system of field equations are found in term of elementary functions. \\
\textbf{Malaver} \citep{Mala13b}: In this paper, Malaver has obtained a realistic model for charged anisotropic matter distribution by considering Van der Waals modified equation of state with polytropic exponent. They made specific choices for metric potential $\lambda=\frac{1}{2}(\log(1+\alpha cr^2)-\log(1+\beta cr^2))$ and electric field intensity $E=\sqrt{\frac{2c\beta cr^2}{(1+\alpha cr^2)^2}}$. For $\beta = 0$, they found uncharged model with polytropic index $\eta = 1, 2$. They have also claimed that some known solutions may be recovered
for specific values of adjustable parameters.\\
\textbf{Malaver $\&$ Kasmaei} \citep{Malaver19}: They have studied the behavior of relativistic compact objects with anisotropic matter distribution considering Van der Waals modified equation of state and a gravitational potential ($\lambda=-\log{(1-\alpha acr^2)}$) that depends on an adjustable parameter $\alpha$. They have generated the new exact solutions of the Einstein-Maxwell system. All the physical variables are written in terms of elementary and polynomial functions. They have obtained expressions for radial pressure, density, and mass of the stellar object physically acceptable with two different values of the adjustable parameter ($\alpha=\frac{1}{2}, 2$). Their results correspond to strange star models for 4U 1820-30 and SAX J 1808.4-3658. These models satisfy all required physical features of a relativistic star.
\clearpage
\section{Solutions when spacetime admit conformal motion}\label{sec4}
In this section, we have considered the solutions obtained by assuming the spacetime in conformal motion. The existence of conformal motion or flatness, in particular, provides an additional relationship between both the metric potential, which again reduces the original problem into the system of four equations for six unknowns. One can obtain the solution by assuming two of the fluid characteristics as known one or imposing additional constraints in field equations.

The complicated nature of the field equations allows the imposition of supplementary constraints like infinitesimal point transformations which lead to conformal motion. These are usually imposed on the system of field equations to make them easily integrable. The possibility of finding physically acceptable solutions with the condition of conformal motion is probed into its dependence crucially on the form of the conformal Killing vector field $\xi$.
To obtain the natural relationship between matter and geometry through the gravitational field equations, inheritance symmetry can be used. The well-known inheritance symmetry is the symmetry under conformal Killing vectors, which is given by,
\begin{eqnarray}
	L_\xi g_{ik} =\psi g_{ik}
	\label{cvk}
\end{eqnarray}
where $L_\xi$ is the Lie derivative of the metric tensor, which describes the interior gravitational field of a compact star, with respect to the vector field $\xi$, and $\psi$ is the conformal factor. It is considered that the vector $\xi$ generates the conformal symmetry and the metric $g$ is conformally mapped onto itself along $\xi$. If $\psi=0$ then (\ref{cvk}) gives the Killing vector. Also, in this case, the underlying spacetime is asymptotically flat, which further implies that the Weyl tensor will also vanish.
Manjonjo et al. \citep{Manjonjo} have shown that the existence of a conformal Killing vector implies a relationship relating the gravitational potentials as follows:
\begin{eqnarray}
	2\nu''+\nu'^2=\lambda'\nu'+\frac{2\nu'-\lambda'}{r}+\frac{4}{r^2}[(1+\mathcal{S})e^\lambda-1]
\end{eqnarray}
where $\mathcal{S}$ is a constant, which is zero for conformally flat space-time.
Imposing this relation provides a systematic method of generating solutions of the Einstein and Einstein-Maxwell systems of field equations.

Several authors have studied the field equations for neutral matter as well as charged matter with conformal symmetry. Solutions generated by this approach are very useful in relativistic astrophysics and can be used to model dense stars. Because of its potential applications in cosmology and astrophysics, most studies in conformal motions have been done in spherically symmetric spacetime. It has been acknowledged that static spherical geometries do admit conformal symmetries which are nonstatic in nature. Conformal motions in static and spherically symmetric spacetime have been substantially studied by Maartens et al. \citep{Maartens95,Maartens96} and Tupper et al. \citep{Tupper12}. Several researchers have used conformal symmetries to model compact fluid spheres in a general relativistic setting. Anisotropic stars admitting conformal motion have been studied by Rahaman et al. \citep{Rahman10,Rahman102}. Relativistic stars admitting conformal motion have been analyzed in \citep{Rahman102}. Herrera et al. \citep{Herrera84} and Herrera and Ponce \citep{Herrera85,Herrera852}, Maartens and Maharaj \citep{Maartens90} and Mak and Harko \citep{Mak} have modeled charged anisotropic fluids in the presence of conformal symmetry. New classes of charged fluid spheres with conformal symmetry and a linear equation of state were found by Esculpi and Aloma \citep{Esculpi10}. Rahaman et al. \citep{Rahman15} and Bhar \citep{bhar14} have described conformal motion in higher dimensional spacetime. Rahaman et al. \citep{Rahman 17} and Shee et al. \citep{Shee} studied anisotropic stars with a nonstatic conformal vector and a selected spacetime geometry. Thus, we can say that the assumption of conformal symmetry has been useful in generating realistic astrophysical models.

In Table \ref{conformal} we have given some metrics which have been used to study exact solutions of the Einstein field equations by considering spacetime in conformal motion.
\begin{sidewaystable}[h]
	\centering
	\caption{\label{conformal} Solutions obtained by considering spacetime admit conformal motion.}
	\tiny	\begin{tabular}{|l|l|}
		\hline
		name [ref.] & metric  \\
		\hline
		R-J-S-C \citep{Rahman10}&$ds^2=A^2r^2dt^2-\frac{B^2}{(1-a)B^2-B^2br^2+\frac{C}{r}}dr^2-r^2d\Omega^2$\\
		\hline
		Bhar \citep{Bhar15}&$ds^2=B^2r^2dt^2-\frac{R^2-r^2K^2}{R^2-r^2}dr^2-r^2d\Omega^2$\\
		\hline
		M-M-R Ia \citep{Matondo}&$ds^2=-A^2 \frac{a^2cr^2}{4K^2}\left[K^2(\sqrt{c}b)^2(1+dcr^2)^{\frac{-bd}{2(d-e)}}(1+ecr^2)^{\frac{be}{2(d-e)}}+(\sqrt{c}b)^{-2}(1+ecr^2)^{\frac{-be}{2(d-e)}}(1+dcr^2)^{\frac{bd}{2(d-e)}}\right]^2dt^2$\\
		&$+\frac{1}{[(1+d cr^2)(1+e cr^2)]^2}dr^2+r^2d\Omega^2$; $b\ne e$.\\
		\hline
		M-M-R Ib \citep{Matondo}&$ds^2=-A^2 \frac{a^2cr^2}{4K^2}\left[ K^2 \left(-\frac{becr^2}{1+ecr^2}\right)^{b/2} \exp\left(\frac{b}{(1+ecr^2)^2}\right) +\left(-\frac{becr^2}{1+ecr^2}\right)^{-b/2} \exp\left(-\frac{b}{(1+ecr^2)^2}\right)\right]^2dt^2+\frac{1}{(1+e cr^2)^4}dr^2+r^2d\Omega^2$\\
		\hline
		M-M-R IIa \citep{Matondo}&$ds^2=-A^2 \frac{a^2cr^2}{4K^2}\left[K^2 (cr^2)^{\frac{b}{2 \alpha \beta}}(\alpha+cr^2)^{\frac{b}{2\alpha(\alpha-\beta)}}(\beta+cr^2)^{-\frac{b}{2\alpha(\alpha-\beta)}}+(cr^2)^{-\frac{b}{2 \alpha \beta}}(\alpha+cr^2)^{\frac{b}{2\alpha(\alpha-\beta)}}(\beta+cr^2)^{-\frac{b}{2\alpha(\alpha-\beta)}}\right]^2dt^2$\\
		&$+\frac{1}{(1+d c^2r^4+e cr^2)^2}dr^2+r^2d\Omega^2$; $\alpha= \frac{e-\sqrt{e^2-4d}}{2d}$, $\beta=\frac{e+\sqrt{e^2-4d}}{2d}$, $e^2> 4d$.\\
		\hline
		M-M-R IIb \citep{Matondo}&$ds^2=-A^2 \frac{a^2}{4K^2}\left[\frac{K^2 (cr^2)^{\frac{b+1}{2}}}{(1+\sqrt{d} cr^2)^{\frac{b}{2}}} \exp \left(\frac{b}{(1+\sqrt{d} cr^2)}\right)+\frac{{(1+\sqrt{d} cr^2)^{\frac{b}{2}}}}{(cr^2)^{\frac{b-1}{2}}} \exp \left(-\frac{b}{(1+\sqrt{d} cr^2)}\right) \right]^2dt^2+\frac{1}{(1+2cr^2\sqrt{d}+d c^2r^4)^2}dr^2+r^2d\Omega^2$\\
		\hline
		M-M-R IIIa \citep{Matondo}&$ds^2=-A^2 \frac{a^2cr^2}{4K^2}\left[\frac{K^2 (cr^2)^{\frac{b}{2 \alpha \beta}} (\beta+cr^2)^{\frac{b}{2\beta(\alpha-\beta)(n\beta-1)}}}{(1+ncr^2)^{\frac{n^2b}{2(n\alpha-1)(n\beta-1)}(\alpha+cr^2)^{\frac{b}{2\alpha(n\alpha-1)(n\beta-1)}}}}+
		\frac{(1+ncr^2)^{\frac{n^2b}{2(n\alpha-1)(n\beta-1)}}(\alpha+cr^2)^{\frac{b}{2\alpha(n\alpha-1)(n\beta-1)}}}{(cr^2)^{\frac{b}{2 \alpha \beta}}(\beta+cr^2)^{\frac{-b}{2\beta(\alpha-\beta)(n\beta-1)}}}\right]^2dt^2$\\
		&$+\frac{1}{[(1+ncr^2)(1+d c^2r^4+e cr^2)]^2}dr^2+r^2d\Omega^2$; $\alpha=\frac{e^2-\sqrt{e^2-4d}}{2d}$, $\beta=\frac{e^2+\sqrt{e^2-4d}}{2d}$, $e^2>4d$.\\
		\hline
		M-M-R IIIb \citep{Matondo}&$ds^2=-A^2 \frac{a^2}{4K^2}\left[\frac{K^2 (cr^2)^{\frac{b+1}{2}} (1+\sqrt{d}cr^2)^{\frac{b({2n\sqrt{d}-d})}{2(\sqrt{d}-n)^2}}}{(1+ncr^2)^{\frac{n^2b}{2(\sqrt{d-n})^2}}} \exp{\left(\frac{b\sqrt{d}}{2(\sqrt{d}-n)(1+\sqrt{d}cr^2)}\right)}+
		\frac{(1+ncr^2)^{\frac{n^2b}{2(\sqrt{d}-n)^2}}} {(cr^2)^{\frac{b-1}{2}} (1+\sqrt{d}cr^2)^{\frac{b({2n\sqrt{d}-d})}{2(\sqrt{d}-n)^2}}}\right]^2dt^2 +$\\
		&$\frac{1}{[(1+ncr^2)(1+d c^2r^4+e cr^2)]^2}dr^2+r^2d\Omega^2$\\
		\hline
		M-M-D \citep{Maurya2019}&$ds^2=\left[(Ar^2+B)\exp{\left(\frac{a(\sqrt{\alpha(K+1)})^{1-n}-br^2}{2b(n-1)}\right)}\right]^2dt^2-\frac{1}{\alpha (K+1)}\left[\frac{\alpha r^2+(\sqrt{\alpha (K+1)}-br^2)^n}{(\sqrt{\alpha (K+1)}-br^2)^n}\right]^2 dr^2-r^2d\Omega^2$; $\alpha=\pm1$,\\
		& $K\ne-1$, $0<n\ne1$\\
		\hline
	\end{tabular}
\end{sidewaystable}
Working Rule:\\
\textbf{Rahaman et al.} \citep{Rahman10}: In this manuscript, the authors have obtained a new class of interior solutions for neutral anisotropic stars in spacetime admitting conformal motion. The Einstein’s field equations in this construction are solved for specific choices of the density function ($\rho=\frac{1}{\kappa}(\frac{a}{r^2}+3b)$). They have analyzed the behavior of the model parameters like radial and transverse pressures, density as well as surface tension and shown that all these parameters are well-behaved. However, the obtained solutions suffer from a central singularity problem, and, they claimed that the model is suitable for the description of the envelope region of the star.\\
\textbf{Bhar} \citep{Bhar15}: In this paper, Bhar has obtained a new model of an anisotropic superdense star, admitting conformal motions in the presence of a quintessence field, characterized by a parameter $\omega_q$ with $-1 < \omega_q <-\frac{1}{3}$. The model has been developed by choosing the Vaidya–Tikekar ansatz \citep{Vaidya}. This model satisfies all the physical requirements. He has analyzed the obtained result analytically as well as using graphical representations. The model is potentially stable and in static equilibrium under anisotropic, gravitational, and hydrostatic forces.\\
\textbf{Matondo et al.} \citep{Matondo}: They have presented exact models for uncharged anisotropic gravitating stars with conformal symmetry. They have used this new relationship between the metric potentials due to conformal motion and considered specific choices for metric potential $\lambda$ to find new classes of exact solutions. They have studied the physical features of the model and demonstrated that the model is well behaved. They have shown that the criteria for stability have been satisfied. They regained masses, radii, and surface redshifts for the compact objects PSR J1614-2230 and SAX J1808.4-3658.\\
\textbf{Maurya et al.} \citep{Maurya2019}: They have generated a new family of exact solutions to the system of field equations with a neutral anisotropic fluid distribution
for a spherically symmetric spacetime that contains a conformal Killing vector. The obtained matter variables and the metric functions are regular at the center and well behaved throughout the interior of the star. They have shown that the solution is physically viable and claimed that it may be utilized to model a compact object. They have shown the physical acceptability of the model through a detailed graphical analysis of the matter variables. The model is shown stable in terms of cracking, the adiabatic index inequality, and, the Harrison–Zeldovich–Novikov stability criterion. In short, they have proved that for an astrophysical application, it is possible to geometrically describe a compact object with conformal symmetry.\\
\clearpage
\section{Solutions obtained by considering spacetime of embedding class one}\label{sec5}
The embedding problem is one of the interesting problems of geometrically significant spacetime. This was first addressed by Schlai \citep{Schlai}. Karmarkar \citep{Karmarkar} derived the condition for embedding 4-dimensional spacetime metric in 5-dimensional Euclidean space, and classified these spacetime as class-1 space-time. spacetime of embedding class one is a special case of the more general class $m$. An $n$-dimensional Riemannian space is called of embedding class $m$ if $m+n$ is the lowest dimension of the flat space in which the given space can be embedded. The resulting mathematical model has proved to be extremely useful in the study of compact stellar objects \citep{MG16,Maurya2016,MG17,MM17,MR17,MGR19,MD19}.
To solve the system of Eqs. (\ref{fe1})–(\ref{fe3}) one can employ the method used by Karmarkar \citep{Karmarkar}, where the obtained solutions are classified as class one spacetime. In this method the Riemann curvature tensor $R_{\alpha\beta\mu\nu}$ satisfies a particular equation that finally links the two metric component $e^\nu$ and $e^\lambda$ in a single equation. The nonzero components of Riemann curvature satisfy the Karmarkar condition
\begin{eqnarray}
	R_{1010}R_{2323}-R_{1212}R_{3030}=R_{1220}R_{1330}
	\label{kc}
\end{eqnarray}
where, $R_{1010}=-e^\nu\left(\frac{\nu''}{2}-\frac{\lambda'\nu'}{4}+\frac{\nu'^2}{4}\right)$, $R_{2323}=-e^{-\lambda}r^2\sin^2{\theta}(e^\lambda-1)$, $R_{1212}=-\frac{r}{2}\lambda'$ and $R_{3030}=-\frac{r}{2}\nu'e^{\nu-\lambda}\sin^2{\theta}$.
However, Pandey and Sharma \citep{pandey} pointed out that the above condition is only a necessary one but it is not sufficient to spacetime become class one. In order to be class one a spacetime must satisfies (\ref{kc}) along with $R_{2323} \ne 0$.
Equation (\ref{kc}) transforms into a differential equation for $\lambda$ and $\nu$:
\begin{eqnarray}
	2\frac{\nu''}{\nu'}+\nu'=\frac{\lambda'e^\lambda}{e^\lambda-1}
	\label{kc1}
\end{eqnarray}
with $e^\lambda \ne 1$. On solving (\ref{kc1}), the Karmarkar condition reduces to
\begin{eqnarray}
	e^{\nu/2}=A+B\int \sqrt{e^\lambda(r)-1} dr
\end{eqnarray}
where $A$ and $B$ are constants of integration. The above expression establishes a relationship between both the metric potentials. Using the Karmarkar condition, the expression
for anisotropy (\ref{fe4}) get transformed into \citep{MG16}
\begin{eqnarray}
	\Delta(r)=\frac{\nu'}{4\kappa e^\lambda}\left[\frac{2}{r}-\frac{\lambda'}{e^\lambda-1}\right]\left[\frac{\nu'e^\nu}{2rA_2^2}-1\right]
\end{eqnarray}
Recently, many authors have also presented anisotropic compact star models in the background of embedding class I space-time \citep{SP161,SPP16,SPP16, Bhar2017,Ortiz,MD19,Prasad19,Bhar2019,Sarkar}. In Table \ref{class1} we have given some metrics which have been used to study exact solutions of the Einstein field equations by considering spacetime of embedding class one.
\begin{sidewaystable}[h]
	\centering
	\caption{\label{class1} Solutions obtained by considering spacetime of embedding class one.}
	\begin{tabular}{|l|l|}
		\hline
		name [ref.] & metric  \\
		\hline
		M-G-S-R \citep{MG16}&$ds^2=\left[A+\frac{B\sqrt{(a-b)(1+br^2)}}{b}\right]^2dt^2-\left[1-\frac{(a-b)r^2}{1+br^2}\right]dr^2-r^2d\Omega^2$, $a\ne b$\\
		\hline
		S-P I \citep{SP161}	&$\left[A+\frac{\sqrt{a}B\log{(br^2+1)}}{2b}\right]^2dt^2-\left[1+{a^2r^2}{(1+br^2)^2}\right]dr^2-r^2d\Omega^2$; $\frac{\sqrt{2}}{a}<\frac{B}{A}\le 2\sqrt{a}$.\\
		\hline
		S-P II \citep{SP16}	&$\left[A+\frac{\sqrt{a}B(br^2+1)^{\frac{n}{2}+1}}{b(n+2)}\right]^2dt^2-\left[1+{a^2r^2}{(1+br^2)^n}\right]dr^2-r^2d\Omega^2$, $n\ne-2$\\
		\hline
		S-P-P \citep{SPP16}&$ds^2=\left[A-\frac{\sqrt{3}B}{\sqrt{a(2-ar^2)}}\right]^2dt^2-\frac{2(1+ar^2)}{2-ar^2}dr^2-r^2d\Omega^2$\\
		\hline
		M-M \citep{MM17}	&$ds^2=A^2[B+\tan^{-1}\sinh(ar^2+b)]dt^2-\frac{1+2cr^2+cosh[2(ar^2+b)]}{1+cosh[2(ar^2+b)}dr^2-r^2d\Omega^2$\\
		\hline
		M-R-G \citep{MR17}&$ds^2=\left[A+B\left(e^{(ar^2+b)}-e^{(ar^2+b)}\right)\right]^2dt^2-\frac{4+Cr^2[e^{(ar^2+b)}-e^{(ar^2+b)}]^2}{4}dr^2+r^2d\Omega^2$; $a,b,C>0$\\
		\hline
		B-S-M \citep{Bhar2017}&$ds^2=\left[A-\frac{aB}{2b(1+br^2)}\right]^2dt^2-\left[1+\frac{a^2r^2}{(1+br^2)^4}\right]dr^2-r^2d\Omega^2$\\
		\hline
		O-M-E-N-D \citep{Ortiz}	&$ds^2=\left[A+\frac{\sqrt{2a}B\cos(br^2+c)F(r)f_1(r)f_2(r)}{b(n+1)\sqrt{1-\sin(br^2+c)}}\right]^2dt^2-\frac{1}{1+ar^2[1+\sin(br^2+c)]^n}dr^2-r^2d\Omega^2$, where,  \\
		&$F(r)=2F\left[\frac{1+n}{2},\frac{1}{2},\frac{3+n}{2}, \sin^2(\frac{2c+\pi+2br^2}{4})\right]$ represents a Gauss hypergeometric function, \\ &$f_1(r)=\left[\frac{\cos(c+br^2)+\sin(c+br^2)}{2}\right]^{-n}$ $\&$ $f_2(r)=[1+\sin(c+br^2)]^n$. \\
		\hline
		M-D-R-K \citep{MD19}	&$ds^2=\left[B(1-Ar^2)\right]^ndt^2-[1+n^2A^2BKr^2(1-Ar^2)^{n-2}]dr^2-r^2d\Omega^2$; $n\le -3$\\
		\hline
		P-K-M-D \citep{Prasad19}	&$ds^2=\left[A+\frac{Ba}{2b}\log{\frac{e^{br^2+c}-1}{e^{br^2+c}+1}}\right]^2dt^2-[1+a^2r^2\cosh^2(br^2+c)]dr^2-r^2d\Omega^2$\\
		\hline
		Bhar \citep{Bhar2019}&$ds^2=\left[A-\frac{B\sqrt{1-A_1r^2}}{\sqrt{A_1}}\right]^2dt^2-[1+a^2r^2(1+b^5r^5)^{-n}]dr^2-r^2d\Omega^2$\\
		\hline
		S-S-S-R \citep{Sarkar}&$ds^2=\frac{1}{4a^2\pi}\left[2aA\sqrt{\pi}+B\sqrt{c}e^{-(1+ar^2)^2}+\sqrt{\pi}(1+ar^2) \{erf[1+ar^2]\}\right]^2dt^2-[1+cr^2 \{erf[1+ar^2]\}^2]dr^2$\\
		&$-r^2d\Omega^2$, where, error function $erf[x]$ is defined as $erf[x]=\frac{1}{\pi}\int_{-x}^{x}e^{-t^2}dt$.\\
		\hline
	\end{tabular}
\end{sidewaystable}

\textbf{Working Rule:}\\
\textbf{Maurya et al.} \citep{MG16}: The authors have generated new uncharged anisotropic compact star models by using the class one condition. They have considered one of the metric functions as $e^\lambda=1+\frac{(a-b)r^2}{1+br^2}$ and constructed the expression for anisotropy factor by using the pressure anisotropy condition. Thereafter they have obtained energy density and pressures. These physical parameters are well behaved inside the star and satisfy all the required physical conditions. Also, all the physical parameters of the models are dependent on the anisotropy factor, which is a very interesting feature. The mass and radius of the compact star models are shown to be compatible with the observational astrophysical compact stellar objects like SAX J1808.4-3658, SAX J1808.4-3658, Her X-1, and RXJ 1856-37.\\
\textbf{Singh $\&$ Pant} \citep{SP161}: In this work, the authors have obtained a family of exact solutions for relativistic uncharged anisotropic stellar objects by considering a 4-D spacetime embedded in a 5-D pseudo-Euclidean space. They have generated the solutions by assuming a new metric potential given by $e^\lambda=1+ar^2(1+br^2)^n$. The resulted solutions are well behaved in all respects and satisfy all energy conditions. The resulting compactness parameter is within the Buchdahl limit. The well-behaved nature of the solutions for a given star candidate is completely dependent on the index $n$. The authors have given a detailed discussion on the solutions for the neutron star XTE J1739-285. For this particular star, the solution is well behaved in all respects for $8 \le n \le 20$. However, the solutions with $n < 8$ possess an increasing trend for the velocity of sound, and the solutions belonging to $n > 20$ violate the causality condition. Also, the well-behaved nature of the solutions for PSR J0348+0432, EXO 1785-248, and Her X-1 are specified by the index n with limits $24 \le n \le 54$, $1.5 \le n \le 4$, and $0.8 \le n \le 2.7$, respectively.\\
\textbf{Singh et al.} \citep{SPP16}: The authors have presented a solution of embedding class I that describes the interior of a spherically symmetric charged anisotropic stellar configuration. The exact analytic solution has been obtained by considering Buchdahl \citep{Buchdahl} type metric potential. Using this solution, the authors have discussed various physical aspects of a compact star. The solution is free from central singularities.  The counter-balancing of different forces acting on the fluid system has been represented by TOV
equation, which signifies that the solution represents a stable hydrostatic equilibrium. The solution also satisfies the energy conditions. The calculated masses and radii of compact stars RX J1856.5-3754, XTE J1739-285, PSR B0943+10, and SAX J1808.4-3658 have been compared with their observational values.\\
\textbf{Maurya $\&$ Maharaj} \citep{MM17}: By analyzing the Karmarkar embedding condition, the authors have generated anisotropic solutions for spherically symmetric spacetime. They have constructed a suitable form of one of the gravitational potentials as $e^\lambda=\frac{1+2cr^2+cosh[2ar^2+b]}{1+cosh[2ar^2+b]}$ to obtain a closed-form solution. The resulting solution is well behaved, which has been utilized to construct realistic static fluid sphere models. The physical validity of the model depends on the parameter values of a, b and c.  Also,  by using observational data set values, they have estimated the masses and radii for compact stars PSR J1903+327, EXO 1785-248, LMC X-4, and 4U 1820-30, which show that this model is suitable for representing fluid spheres to a very good degree of accuracy. \\
\textbf{Maurya et al.} \citep{MR17}: The authors have presented a class of new solutions describing neutral anisotropic stellar configuration satisfying Karmarkar's condition. This compact star model is physically well-behaved and meets all the physical requirements for a stable configuration in hydrostatic equilibrium. This model is compatible with compact stars 4U1608-52 and Vela X-1 to a very good approximation.\\
\textbf{Bhar et al.} \citep{Bhar2017}: In this paper, the authors have constructed a new relativistic anisotropic compact star model having a spherically symmetric metric of embedding class one. They have assumed $e^\lambda=1+\frac{a^2r^2}{(1+br^2)^4}$ and solved the Einstein’s field equations with the help of Karmarkar condition for an anisotropic matter distribution. A detailed discussion of the physical properties of the model (pressure, density, mass function, surface red-shift, gravitational redshift) are investigated and the stability of the stellar configuration are given. The model is free from central singularities, satisfies all the energy conditions, and satisfy the Harrison–Zeldovich–Novikov stability criterion for $0 \le \rho_c \le 4.04 × 1017 \ g/cm^3 $, i.e., the region is stable. However, for $\rho_c \ge 4.04 × 1017 \ g/cm^3$, the region is unstable.\\
\textbf{Ortiz et al.} \citep{Ortiz}: In this work, the authors have presented a completely new class of well-behaved solutions to field equations describing uncharged anisotropic matter distribution in the embedding class one spacetime framework using Karmarkar’s condition. They have performed this analysis by proposing a new metric potential $e^\lambda=1+ar^2[1+sin(br^2+c)]^n$ which plays a key role in generating a realistic model. They have represented the physical features of the solution analytically as well as graphically for the strange star  SAX J1808.4-3658, with $0.5\le n\le 3.4$. The resulted solutions are free from singularities, satisfy Abreu’s criterion \citep{Abreu}, causality condition as well as relativistic adiabatic index, and exhibit well-behaved nature. The equilibrium condition, as well as the energy conditions, are well-defined. These features confirm the physical acceptability of the model. The authors have shown that the obtained moment of inertia is as per the Bejger-Haensel concept \citep{Bejger}, and could provide a precise tool to the matching rigidity of the state equation due to different values of $n$.\\
\textbf{Maurya et al.} \citep{MD19}: In this manuscript, the authors have discussed a generalized model for neutral anisotropic compact stars in the presence of highly dense and ultra-relativistic matter distribution. After embedding the 4D Riemannian space locally and isometrically into a 5D pseudo-Euclidean space, they have solved the field equations by employing a class of physically acceptable metric functions $e^\nu=B(1-Ar^2)^{n}$, $n\le-3$. The same form of metric potential was previously used by Maurya et al. \citep{MGR19} for $n>2$, and Maurya and Govender \citep{MG17} for $n\le-1$ to model charged isotropic compact stars arising from the Karmarkar condition. In the obtained model, the anisotropy is zero at the center and maximum at the surface. The usual energy conditions are satisfied and that the compact structures are shown to be stable, based on several criteria, viz., the equilibrium of forces, Herrera cracking concept, and adiabatic index. This stellar model also satisfies the Buchdahl condition. Finally, the values of the numerous constants and physical parameters are determined for the compact star LMC X-4, which has been considered as a representative of the compact stars to present the analysis of the obtained results. The authors have claimed that this model can justify most of the compact stars including white dwarfs and ultra-dense compact stars for suitable parametric values of $n$.\\
\textbf{Prasad et al.} \citep{Prasad19}: In this paper, the authors have presented a family of anisotropic compact stars by solving Einstein’s field equations. The field equations have been solved by a particular choice for the metric potential $\lambda$ as $e^\lambda=1+a^2r^2 \csc h^2[br^2+c]$ and embedding class one condition. The physical analysis of this model indicates that the obtained model for relativistic anisotropic stellar structure is physically viable for a compact star whose energy density of the order 1015 g/cm3. They have explored the hydrostatic equilibrium and the stability of the compact stars  Vela X-1, 4U 1820-30, EXO 1785-248, PSR J1903+327, 4U 1538-52, SMC X-1, Her X-1, PSR J1614-2230, SAX J1808.4-3658, 4U 1608-52, LMC X-4, RX J1856-37, and Cen X-3. \\
\textbf{Bhar} \citep{Bhar2019}: In this article, finite, regular, and exact class one solutions of Einstein’s field equations have been presented. To generate the model, the authors have introduced a completely new metric potential as $1+cr^2 \{erf[1+ar^2]\}^2$, with $c>0, a\ne0$ and $erf[x]=\frac{1}{\pi}\int_{-x}^{x}e^{-t^2}dt$. Various physical aspects of the model such as energy density, pressure, anisotropy,  equilibrium, energy conditions, stability, mass, compactness parameter, surface red-shift, gravitational red-shift, and their graphical representations have been investigated, which ensure that the obtained solutions are well-behaved and therefore, represent physically acceptable models for anisotropic fluid spheres. The models has also satisfied the causality condition. It is shown to be stable by satisfying the TOV equation, Herrera cracking concept, Adiabatic index bound, and Harrison–Zeldovich–Novikov condition. The solutions are representing Cen X-3, EXO 1785-248, Vela X-1, and LMC X-4. From the solutions, the M–R graph has been generated and it matches the ranges of masses and radii for the considered compact stars. In this work, the authors have also estimated the approximate moment of inertia for the mentioned compact stars.

\clearpage
\section{Conclusion}\label{sec6}
Stellar models consisting of spherically symmetric distribution of charged as well as uncharged anisotropic matter in gravitational fields have been extensively considered in the frame of general relativity. To investigate these structures, we generate exact models via solving the Einstein-Maxwell (or Einstein) system of equations. Apart from the imposition of symmetry, the fall-off behavior of the pressure and density is considered to obtain a physical model. The vanishing radial pressure at the boundary has received considerable recognition. The produced values of redshifts, luminosity, and mass for the stars in the presence of charged matter are different from the case of neutral matter. To solve the system of equations governing the gravitational and thermodynamical behavior of bounded compact objects, several techniques were employed by researchers such as ad-hoc assumptions of the gravitational potentials, specific choices of pressure, density, or the anisotropy factor, imposition of the equation of state, etc. Some of them have assumed the spacetime to be in conformal motion. A lot of models have been generated by considering spacetime of embedding class one, obeying the Karmarkar condition. In this article, we have characterized the interior solutions of the Einstein-Maxwell/Einstein system of equations for the spherically symmetric distribution of anisotropic matter based on the technique used to generate it.

There are two main ways to generate an anisotropic compact star model irrespective of it being charged or uncharged, where the neutral case can be considered as the charged one, with a given charge function zero. One can consider that the components of the energy-momentum tensor are mutually independent. Consecutively the system of field equation is solved by assuming any three of the six fluid characteristics as known one. Alternatively, we can impose some additional constraints on the system of field equations, assumming an equation of state, spacetime admitting conformal motion, or higher-dimensional spacetime (spacetime of embedding class one in particular). These constraints establish relation between the fluid characteristics and thus, these parameters remain no more independent to each other. 

Generally, in the second approach after imposing one constraint on the system, two more fluid characteristics should be taken as given one, so that the resulting system becomes solvable. However, there are solutions present in the literature in which more than one constraint have been imposed. One can assume the equation of state along with the assumption that spacetime is of embedding class one, or the equation of state in the spacetime assenting conformal motion \citep{Mak,Esculpi10}, or the spacetime of embedding class one validating conformal motion. Futhermore, one fluid characteristic must be known in all mentioned cases to obtain the solutions. To solve the Einstein-Maxwell/Einstein system of equations one can also consider the spacetime of embedding class one, admitting conformal motion and an equation of state to exist. In this approach, no additional information is required to generate the model.

To sum it up, we have outlined the different ways to solve the system of Einstein-Maxwell (or Einstein) equations for compact stars having anisotropic matter distribution. Our goal was to present a comprehensive but still thorough review of anisotropic fluid solutions in general relativity to provide a basis for beginners and a reference guide for the keen ones. Even though we have tried to cover all aspects, no review can match the actual study of the literature itself. It is possible that certain readings might have been omitted, or analyzed less rigorously and, therefore, the reader can resort to the original contents if necessary.

\section*{Declarations}

\begin{itemize}
	\item Funding: Science and Engineering Research Board (SERB), DST, New Delhi.
	\item Conflict of interest: The authors declare that they have no conflict of interest.
	\item Ethics approval: The authors have adhered to the accepted ethical standards of a genuine research study.
	\item Consent to participate: The authors give their concent to participate in this work.
	\item Consent for publication: The authors give their consent for publication.
	\item Availability of data and materials: Not applicable
	\item Code availability: Not applicable
	\item Authors' contributions: J. Kumar: study concept, critical revision of the paper, study supervision. 
	P. Bharti: literature survey, designing and drafting of the paper. All the authors have read and approved the final version of paper.
\end{itemize}

\end{document}